\documentstyle[12pt,epsfig]{article}

\begin{document}
\sloppy
\thispagestyle{empty}

\mbox{}
\vspace*{\fill}
\begin{center}
{\LARGE\bf The cross-section of muon-nuclear } \\

\vspace{2mm}
{\LARGE\bf inelastic interaction}\\

\vspace{2em}
\large
A.V.Butkevich\footnote{E-mail address: butkevic@al20.inr.troitsk.ru} and 
S.P.Mikheyev\footnote{E-mail address: mikheyev@psbai10.inr.ruhep.ru}
\\
\vspace{2em}
{\small\it Institute for Nuclear Research of Russian Academy of Science,}
 \\
{\small\it 60th October Anniversary prospect, 7a, Moscow 117312,
 Russia}\\
\end{center}
\vspace*{\fill}
\begin{abstract}
\noindent
It is shown that the combination of the structure functions $F_2$ predicted by
the CKMT model at low and moderate values of $Q^2$ and MRS99 parton
distribution functions at high $Q^2$ gives a good description of the data 
over complete measured region of $x$ and $Q^2$. Using these structure 
functions the main characteristics of the muon-nucleus inelastic scattering
 are calculated. Nuclear effects and contributions of the neutral current
 and $\gamma -Z$ interference are taken into account.
\end{abstract}
\vspace*{\fill}
\newpage
%
\section{Introduction}
\label{sect1}
Muon inelastic scattering off nuclei plays an important role in muon
propagation through matter. In this process muon may loss a significant part 
of its energy and can be scattered at large angle.
Therefore, muon-nucleus inelastic scattering is of interest for numerous
applications related to muon transport in matter, in particular, for 
calculations of muon intensity at large depth of matter, muon-induced hadron
 flux underground, background produced by  atmospheric muons in underground 
neutrino experiments, etc.

Several models [1-4] have been developed to describe the muon-nuclear inelastic
interactions, nonetheless uncertainties of this process are much larger 
 than for purely electromagnetic interactions. The reason is that the bulk of 
this process is characterized by low squared four-momentum transfer $Q^2$. 
The smallness of the $Q^2$ does not allow to use of the perturbative QCD 
(pQCD) for calculation of nuclear structure function (SF) 
and phenomenological models such as the Regge or General Vector Dominance 
Model (GVDM) have to be used. The parameterization of the nucleon SF obtained 
in these models, depends on free parameters which can be determined from a fit
of experimental data and can be applied in the range of $Q^2 \le 1-3$
 GeV$^2$. This range is often referred to as photo-production. However, they 
fail to describe deep inelastic scattering (DIS) data at high $Q^2$. 
 The pQCD (NLO QCD) gives a good description of the structure functions at 
$Q^2\ge 3$ GeV$^2$. So, a model to combine various aspects of these approaches 
is needed to describe the $Q^2$ behavior of nucleon SFs over complete range 
from photo-production to DIS.

The widely used approximation [4] of the muon photo-nuclear cross-section
was obtained twenty years ago in the framework of GVDM. Experimental data 
in the region of $Q^2\le 100$ GeV$^2$ and $x\ge 0.01$ were used for
determing the parameters. Recently, precise data [5] on SF in wide ranges
of $Q^2$ ($0.06\le Q^2\le 10^4$ GeV$^2$) and $x$ ($10^{-6}\le x\le 0.98$) 
have been obtained and new nuclear effects (anti-shadowing and EMC-effect) 
were observed.
  
The main goal of this work is the calculation of the
muon-nucleus inelastic cross-section, based on the modern nucleon SF and on 
the present knowledge of nuclear effects.

The paper is organized as follows. In Sec.2 we give the general
relations and definitions used for description of neutral current charged
lepton-nucleon scattering. The procedure of the calculation of nucleon SF,
 using CKMT Regge model CKMT [6] and MRS99 parton distribution function
(PDF) [7] is described in Sec.3. In Sec.4, nuclear effects and their 
parameterization are described. The total cross-section,
the muon energy losses, and the angular distributions of scattered muons are
given in Sec.5. In the conclusion we summarize the main results of the paper.
\section{Neutral current charged lepton-nucleon scattering cross-section}
\label{sect2}
The cross-section of neutral current charged lepton scattering off nucleon 
\begin{equation}
\l(k) + N(p) \to l(k') + X(p')               
\end{equation} 
is given by the sum of contributions from the processes as shown in Fig.1
Here $k(E,\bar k)$ and $k'(E',\bar k')$ are initial and final 
lepton four-momenta,
$q=k-k'$ is the virtual photon and Z-boson momentum, $p$ and $p'$ are 
the initial nucleon momentum and the total momentum of the final hadrons $X$,
respectively. This process can be described by the four-momentum transfered 
$Q^2$, Bjorken $x$ and lepton's energy loss $\nu$ (or inelasticity $y$) 
defined as
\begin{eqnarray}
Q^2=-& &q^2=(k-k')^2, ~~~~  x=\frac{Q^2}{2pq}, 
\nonumber\\ 
& & \nu=\frac{q \cdot p}{M},~~~~~~~~~~y=\frac{p \cdot q}{p \cdot k}.   
\end{eqnarray}
In laboratory system
\begin{eqnarray}
& & Q^2=2(EE'-\bar k\bar k')-2m^2, ~~~~~~x=\frac{Q^2}{2M\nu},
\nonumber\\ 
& & \nu=E-E',~~~~~ y=\frac{\nu}{E},                
\end{eqnarray} 
where $M$ and $m$ are nucleon and lepton masses, respectively.

The general form of the differential cross-section for scattering of charge 
non-polarized lepton on non-polarized nucleon, summed over the final 
lepton polarizations can be expressed as:
\begin{eqnarray}
\frac{d^2\sigma^{l^{-},l^{+}}}{d\nu dQ^2}=\frac{2\pi\alpha^2}{Q^4E^2}
 &\Big[& E^{l^{-},l^{+}}(x,Q^2)-I^{l^{-},l^{+}}(x,Q^2)+
\nonumber\\
   && Z^{l^{-},l^{+}}(x,Q^2)\Big],                  
\end{eqnarray}
where
\begin{equation}
E^{l^{-},l^{+}}=2xF_1^{el}(x,Q^2)Y_1 + F_2^{el}(x,Q^2)Y_2,         
\end{equation}
\begin{eqnarray}
I^{l^{-},l^{+}}=P_z&\bigg\{& g_V\Big(2xF_1^I(x,Q^2)Y_1+
      F_2^I(x,Q^2)Y_2\Big) \mp 
\nonumber\\
   && g_AxF_3^I(x,Q^2)Y_3\bigg\},       
\end{eqnarray}
\begin{eqnarray}
Z^{l^{-},l^{+}}&=&P_z^2\bigg\{(g_V^2+g_A^2)\Big(2xF_1^Z(x,Q^2)Y_1+
F_2^Z(x,Q^2)Y_2\Big)
\nonumber\\
   &&\mp 2g_Vg_AxF_3^Z(x,Q^2)Y_3 \bigg\},             
\end{eqnarray}
and
\begin{eqnarray}
&&Y_1=\Big(Q^2-2m^2\Big)\frac{\nu}{Q^2},~~~~Y_2=\Big[2E(E-\nu)-\frac{Q^2}{2},
\Big]\frac{1}{\nu}
\nonumber\\
&&Y_3=\Big(E-\nu\Big).                  
\end{eqnarray}
Here the term $P_Z$ accounts for the Z$^0$ propagator is
\begin{equation}
P_z=\frac{G}{\sqrt{2}}~\frac{Q^2}{2\pi\alpha}~\frac{m_z^2}{Q^2+m_z^2},  
\end{equation}
where $G/\sqrt{2}$ is the Fermi constant, $\alpha=1/137$ is the fine structure
 constant, and  $m_z$ is Z-boson mass. The constants of lepton weak coupling
$g_V$ and $g_A$ are
\begin{equation}
g_V=-\frac{1}{2}+2\sin^2\theta_W~~~~~~~~ g_A=-\frac{1}{2},   
\end{equation}
where $\theta_W$ is Weinberg angle.

The functions $F_i^{el,Z}$ are the electromagnetic ($\gamma$-exchange) and 
neutral current ($Z$-exchange) structure functions, respectively. The functions
$F_i^I$ correspond to decomposition over invariant functions of the tensor
($\gamma$-$Z$ interference)
\begin{eqnarray}
W^I&&\approx\sum\bigg\{\Big<{p'}\mid{J^{el}_\alpha}\mid{p}\Big>
\Big<{p}\mid{J^{Z}_\beta}\mid{p'}\Big>+
\nonumber\\
&&\Big<{p'}\mid{J^{Z}_\alpha}\mid{p}\Big>             
\Big<{p}\mid{J^{el}_\beta}\mid{p'}\Big>\bigg\}\delta(p'-p-q).
\end{eqnarray}
The upper sign in Eq.(6) and Eq.(7) corresponds to lepton scattering ($e^{-},
\mu^{-}$), and the lower sign is for antilepton($e^{+},\mu^{+}$) scattering.

The term proportional to the function $F_3^I$, is due to interference 
between the electromagnetic scattering amplitude and the axial-vector current
 weak interaction amplitude. The amplitudes have opposite $C$ parities, so 
the corresponding terms have opposite sings for lepton and antilepton 
scattering. At low $Q^2$ the $\gamma-Z$ interference term is much smaller then
the $\gamma$-exchange one, but it increases linearly with $Q^2$ (Eq.(9)) and
becomes comparable with the $\gamma$-exchange term at $Q^2\approx$10$^3$ 
GeV$^2$.

In terms of parton distributions in LO-approximation, the SFs can be written as
\begin{equation}
F^{el}_{1}=\frac{1}{2}\sum{e^2_q}\left(f_q+f_{\bar{q}}\right),   
\end{equation}
\begin{equation}
F^{el}_{2}=x\sum{e^2_q}\left(f_q+f_{\bar{q}}\right),             
\end{equation}
\begin{equation}                                                
F^{Z}_{1}=\frac{1}{2}\sum\left(v^2_q+a^2_q\right)\left(f_q+f_{\bar{q}}\right),
\end{equation}
\begin{equation}
F^{Z}_{2}=x\sum\left(v^2_q+a^2_q\right)\left(f_q+f_{\bar{q}}\right),  
\end{equation}
\begin{equation}
F^{Z}_{3}=2\sum{v_qa_q}\left(f_q-f_{\bar{q}}\right),                  
\end{equation}
and
\begin{equation}
F^{I}_{1}=\sum{e_qv_q}\left(f_q+f_{\bar{q}}\right),        
\end{equation}
\begin{equation}
F^{I}_{2}=2x\sum{e_qv_q}\left(f_q+f_{\bar{q}}\right),                 
\end{equation}
\begin{equation}
F^{I}_{3}=2\sum{e_qa_q}\left(f_q-f_{\bar{q}}\right).                  
\end{equation}
Here $f_q$ and $f_{\bar{q}}$ are parton distribution functions in the proton, 
$e_q$, $v_q$ and $a_q$ are the charge, vector and axial-vector week couplings 
of quarks. For up-quarks (u, c, t) they are
\begin{equation}
e_{u,c,t}=\frac23~~~~v_{u,c,t}=\frac12-\frac43\sin^2{\theta_W}~~~~
a_{u,c,t}=\frac12                                                    
\end{equation}
and for down-quarks (d,s,b)
\begin{equation}
e_{d,s,b}=-\frac13~~~~v_{d,s,b}=-\frac12+\frac23\sin^2{\theta_W}~~~~
a_{d,s,b}=-\frac12                                                    
\end{equation}
It is seen from Eqs.(4) and (9) that the main contribution to the total 
cross-section is due to photo-production (low $Q^2$ process), however at fixed 
outgoing muon energy the large scattering angle corresponds to high $Q^2$
\begin{equation}
\cos\theta=\left(EE'-Q^2/2-m^2\right)/\mid{k}\mid\mid{k'}\mid.                           
\end{equation}
So, for calculation of muon scattering at large angles it is necessary to 
know the behavior of the nucleon SFs in the wide range of 
$Q^2\approx 0.01 - 10^6$ GeV$^2$.
\section{Low and high-$Q^2$ approximation of the nucleon structure functions}
\label{sect3}
At high-$Q^2$ the QCD predictions for the nucleon SFs are obtained by solving
the DGLAP evolution at NLO approximation in $\overline{MS}$ or DIS schemes.
These equations yield the parton distribution functions at all values of 
$Q^2$ provided that the PDF are given as a function of $x$ at some input scale 
$Q^2_0$=1.2-5 GeV$^2$. The latest global fits performed by several group 
(MRS99 [7], GRV98 [8], and CTEQ5 [9]) give a good description of the 
experimental data. At $Q^2$ below $Q_0^2$  perturbative QCD fails to describe 
data and phenomenological non-perturbative (GVDM or Regge) models are 
required. A considerable number of non-perturbative models have been 
developed [10]-[12] recently. These models  
predict a correct limit of $F_2$ at $Q^2$=0 and give a good description of
 the SFs at low and medium $Q^2$. Thus, neither the non-perturbative 
approaches nor pQCD can be expected to describe the $Q^2$ behavior of the SFs 
over the complete the range from photo-production to DIS. A number of models 
combining QCD and phenomenological  approaches have been developed to 
describe data in the transition region of $Q^2$ (see review [13]). In this 
paper we use the CKMT model [6] at low and moderate $Q^2$ and the MRS99 fit 
of PDF [7] at high-$Q^2$.

The CKMT model proposes the following parameterization of the proton 
structure function $F^p_2$
\begin{equation}
F^p_2(x,Q^2)=F_S^p(x,Q^2)+F_{NS}^p(x,Q^2)                     
\end{equation}
The singlet term
\begin{equation}
F^p_S(x,Q^2)=A_Sx^{-\Delta(Q^2)}\left(1-x\right)^{n(Q^2)+4}
\left(\frac{Q^2}{Q^2+a}\right)^{1+\Delta(Q^2)}                     
\end{equation}
corresponds to the Pomeron contribution which determines the small-$x$
behavior of sea quarks and gluons. The dependence of effective intercept
of the Pomeron, $\Delta$ on $Q^2$ is parametrized as
\begin{equation}
\Delta(Q^2)=\Delta_0\left(1+\frac{2Q^2}{Q^2+d}\right).               
\end{equation}
The $x\to1$ behavior of $F_S(x,Q^2)$ is determined by the function
\begin{equation}
n(Q^2)=\frac32\left(1+\frac{Q^2}{Q^2+c}\right).               
\end{equation}
The parameterization for the non-singlet term which corresponds to the 
secondary ($f$,$A_2$) reggion (valent quark) contribution, is
\begin{equation}
F^p_{NS}(x,Q^2)=Bx^{(1-\alpha_R)}\left(1-x\right)^{n(Q^2)}
\left(\frac{Q^2}{Q^2+b}\right)^{\alpha_R},                     
\end{equation}
where behavior at  $x\to0$ is determined by the secondary reggion intercept
$\alpha_R$. The valence quark distribution can be separated into the
contributions of the u and d valence quarks by replacing
\begin{equation}
F^p_{NS}(x,Q^2)=xU_V(x,Q^2)+xD_V(x,Q^2),                     
\end{equation}
where
\begin{equation}
xU_V(x,Q^2)=B_ux^{(1-\alpha_R)}\left(1-x\right)^{n(Q^2)}
\left(\frac{Q^2}{Q^2+b}\right)^{\alpha_R}                     
\end{equation}
\begin{equation}
xD_V(x,Q^2)=B_dx^{(1-\alpha_R)}\left(1-x\right)^{n(Q^2)+1}
\left(\frac{Q^2}{Q^2+b}\right)^{\alpha_R}.                     
\end{equation}
Normalization conditions for valence quarks in proton
\begin{eqnarray}
& &\int\limits_0^1\frac1x\left[xU_V(x,Q^2)\right]dx=2e^2_u
\nonumber\\
& &\int\limits_0^1\frac1x\left[xD_V(x,Q^2)\right]dx=e^2_d         
\end{eqnarray}
fix the values of parameters $B_u$ and $B_d$ at $Q^2=Q_0^2$.

The limit of $Q^2=0$ corresponds to interaction of real photons.
The total cross-section for real photons can be written as 
\begin{equation}
\sigma^{tot}_{{\gamma}p}(\nu)=\left[\frac{4\pi^2\alpha}{Q^2}F_2(x,Q^2)\right]
_{Q^2=0}.
\end{equation}                                    
One can see from Eqs.(23),(27), and (28) that $F_2\approx Q^2$ at $Q^2\to0$ 
and fixed $\nu$. 
Thus, the parameterization 
\begin{eqnarray}
\sigma^{tot}_{{\gamma}p}(\nu)=4\pi^2\alpha&\Big[&A_sa^{-1-\Delta_0}(2M\nu)^
{\Delta_0}+
\nonumber\\
&&(B_u+B_d)b^{-\alpha_R}(2M\nu)^{\alpha_R-1}\Big]      
\end{eqnarray}                                    
takes place in the CKMT model.

In this way we find parameterizations of both $F_2^p$ and $\gamma\-p$ 
cross-section with 7 free parameters: $a, b, c, d, \Delta_0, \alpha_R$, 
and $A_S$. To determine the parameters we have made a joint fit of
the ${\sigma}_{tot}^{\gamma p}$ data and NMC, E665, SLAC, ZEUS, and H1 data
on the proton SF $F_2$ in the region $0.11\le Q^2\le5.5$ GeV$^2$ and 
$10^{-6}\le{x}\le 0.98$ [5]. As initial condition for the values of 
different parameters, we used those obtained in the previous fit in Ref.[6]. 
A global fit results in the following values of parameters (all dimensional 
parameters are in GeV$^2$): $a$=0.2513, $b$=0.6186, $c$=3.0292, $d$=1.4817,
$\Delta_0$=0.0988, $\alpha_R$=0.4056, and $A_S$=0.12. The values of the 
parameters $B_u=1.2437$ and $B_d=0.1853$, were determined from normalization 
conditions for valence quarks (at $Q^2_0$=2 GeV$^2$).
The quality of the description of all experimental data is very good and  
$\chi^2/d.o.f.$=754.8/600, where only
the statistical errors have been used. Recently, a modified version of the CKMT
model used new data on $F_2^p$ at low $Q^2$ has been published [14]. The 
values of the main parameters are in a good agreement with those, obtained in 
the present work.
 
So, for calculation of $F_2$ in the entire region of $Q^2$ we use the CKMT 
model at
$Q^2\le5$ GeV$^2$, the MRS99 PDF at $Q^2\ge6$ GeV$^2$ and a linear fit between
$F_2^p$(CKMT) and $F_2^p$(MRS99) in the region $5\le{Q^2}\le6$ GeV$^2$. The
result of the fit of $F_2^p$ and $\sigma^{tot}_{\gamma p}$ is shown 
in figures 2 ($F_2^p$ vs. $x$ for different value of $Q^2$) and 3 ($F_2^p$ vs. 
$Q^2$ for different value of $x$) along with the experimental data [5]. The 
cross-section  $\sigma^{\gamma p}_{tot}$ as a function of $W^2=M^2+2M\nu-Q^2$ 
is shown in Fig.4 (the data from Refs.[15],[16]). 

A good description of experimental data is obtained at all $x$ and $Q^2$
values. It should be noted that:

a) the recent ZEUS BPT97 data [17] were not included in our fit, 
but are in agreement with the CKMT model prediction at $Q^2\le0.1$ GeV$^2$,

b) the rise of $F_2^p$ at low $x$ and low $Q^2$ is well described by the CKMT
model with the slope $\Delta_0$=0.0988 while the experimental value is 
0.102$\pm$0.07 [17],

c) the $\sigma^{tot}_{\gamma p}$ values found by the ZEUS collaboration are 
the result of a phenomenologically motivated extrapolation.
 
In figures 2-4 the SF $F_2^p$ and $\sigma_{tot}^{\gamma p}$
that have been obtained by Bezrukov and Bugaev [4] and used for  
calculation of the muon photo-nuclear cross-section are shown also. 
At $x<10^{-3}$ the SFs rise slower than the present data indicate.
An the other hand in the region $x>0.01$ and $Q^2>5$ GeV$^2$ the SFs
are overestimated. 

The CKMT parameterization gives separate contributions of valence quarks,
 sea quarks, and gluons. We used this peculiarity for parameterization of 
the neutron SF $F_2^n$, that can be extracted from the deuteron $F_2^d$ and 
the proton $F_2^p$ data, using the relation

\begin{equation}
F_2^d=\frac12\left[F_2^p(x)+F_2^n(x)\right],      
\end{equation}                                    
and the Gottfried sum rule
\begin{equation}
S_G=\int\limits_0^1\left(F_2^p-F_2^n\right)\frac{dx}{x}=
\frac13\int\limits_0^1(d_V-u_V)dx-\frac23\int\limits_0^1(\bar{d}-
\bar{u})dx.                                     
\end{equation}                                    
In the case of SU(2) symmetric sea, PDF of $\bar{d}$ equals to=$\bar{u}$ and 
therefore $S_G$=1/3. However, the NMC collaboration [5]
 gives $S_G$=0.235$\pm$0.026 at $Q^2=4$ GeV$^2$, i.e. significantly below 1/3 
and shows that $F_2^p(x)-F_2^n(x)\to 0$ and $F_2^n(x)/F_2^p(x)\to 1$ at 
$x\to 0$. Taking into
account these results the singlet term of the $F_2^n$ has to be modified.
Because of the isotopical invariance of strong interaction the non-singlet term
$F_{NS}^n$ is
\begin{equation}
F_{NS}^n(x,Q^2)=\frac14xU_V(x,Q^2)+4xD_V(x,Q^2)         
\end{equation}
where $xU_V(x,Q^2)$ and $xD_V(x,Q^2)$ are given by Eq.(29) and 
Eq.(30). The singlet term 
\begin{equation}
F^n_S(x,Q^2)=A_Sx^{-\Delta(Q^2)}\left(1-x\right)^{n(Q^2)+\tau}
\left(\frac{Q^2}{Q^2+a}\right)^{1+\Delta(Q^2)}                     
\end{equation}
has an additional free parameter $\tau$. The value of this parameter  
obtained from fit of the $F_2^d$ data [5] in the region $Q^2\le 5.$ GeV$^2$
(all other parameters were fixed by the fit of the $F_2^p$ and 
$\sigma^{tot}_{\gamma p}$) is $\tau=1.8152$. The quality of the description 
of data is good with value of $\chi^2/d.o.f=611.1/453$ for the SF $F_2^d$ data 
and $\chi^2/d.o.f=452.8/380$ for the $F_2^n/F_2^p$ data, where only 
statistical errors have been used.

For calculation of the SF $F_2^n$ in the entire region of $Q^2$ we used 
the approximation of Eq.(37) and Eq.(38) at $Q^2\le5.$ GeV$^2$, the MRS99 PDF 
at $Q^2\ge6$ GeV$^2$ and linearly fit between $F_2^n$(CKMT) and $F_2^n$(MRS) 
in the transition region $5<Q^2<6$ GeV$^2$. The SF $F_2^d$ (Fig.5) and 
$F_2^n/F_2^p$ (Fig.6) as function of $Q^2$ for different values of $x$ are 
shown with experimental data. Fig.7 shows the $F_2^p-F_2^n$ vs. $x$ at 
$Q^2=4$ GeV$^2$. The calculations are in agreement with the NMC data [18].

For calculation of the cross-section of lepton-nucleon scattering it is 
necessary to know the behavior of SF $2xF_1$ in a wide range of $Q^2$ and
$x$ also. This SF can be expressed using the longitudinal SF $F_L$ 
\begin{equation}
F_L=\left(1+\frac{4M^2x^2}{Q^2}\right)F_2-2xF_1.           
\end{equation}
Then
\begin{equation}
2xF_1=\frac{1}{1+R}\left(1+\frac{4M^2x^2}{Q^2}\right)F_2,           
\end{equation}
where
\begin{equation}
R=\frac{F_L}{\left(1+4M^2x^2/Q^2\right)F_2-F_L}.           
\end{equation}

Perturbative QCD describes reasonably well the available data 
 on the ratio $R(x,Q^2)$ at large values of $Q^2$ and very little is known 
about possible extrapolations towards the region of low $Q^2$. In the limit 
of $Q^2\to 0$ the SF $F_L$ has to vanish as $Q^4$ (for fixed $\nu$ ) and 
therefore $R\approx Q^2$. 
At $x<0.01$ and $Q^2<0.5$ GeV$^2$ the experimental results are poor. 
 Data show a small value of $R$ at moderate values of $x$ and a possible 
increase of $R$ as $x$ decreases. The data come from experiments carried out 
on different targets and the differences $R^A-R^p$ are consistent with zero 
and do not exhibit any significant dependence on $x$ [19]. 

Data can be fitted by a parameterization $R$(SLAC98) [20]. However, this fit 
should not be used at $Q^2<0.35$ GeV$^2$. In this region we used the GVDM 
asymptotic of $R$ at $Q^2\to 0$ given by [21]
\[
R_{GVMD}(Q^2,x)\approx \frac{Q^2}{Q^2+m_{\rho}^2},
\]
where $m_{\rho}=0.77$ GeV is the $\rho$ meson mass. Thus, at $Q^2>Q^2_0=
1.4$ GeV$^2$ the function $R(x,Q^2)$ is calculated as follows: 
\begin{equation}
R(x,Q^2)=\left\{\begin{array}{l}
R(MRS99)\mbox{~~~~at}~~~~ x< 10^{-3}\\               
R(SLAC98)\mbox{~~~at}~~~~ x\ge 5.10^{-3}\\
\end{array}
\right.
\end{equation}
In the region $10^{-3}<x<5\cdot 10^{-3}$ a linear fit between 
$R$(MRS99, $x=10^{-3}$) and $R$(SLAC98, $x=5\cdot 10^{-3}$) is used. 
At $Q^2<Q^2_0$
\begin{equation}
R(x,Q^2)=R_{GVMD}(x,Q^2)=C(x)\frac{Q^2}{Q^2+m^2_{\rho}},      
\end{equation}
where the function $C(x)$ is determined by normalization condition at $Q_0^2$
\begin{equation}
R_{GVMD}(x,Q^2_0)=R(x,Q^2_0),                                 
\end{equation} 
and  the function $R(x,Q^2_0)$ is calculated using Eq.(40). 
Fig.8 shows the experimental values of $R$ as a function of $Q^2$ in four 
ranges of $x$ along with the result of parameterization Eq.(41)-Eq.(42). In the
region of low $Q^2<Q^2_0$ it decreases with $Q^2$ at all values $x$, but the
dependence on $x$ is not strong (Fig.9).  
However, extrapolation of $R$ outside the kinematical range
of data: namely at $Q^2\to0$ and $x\to0$, based on the presently available
data, is a rather delicate problem.

In Figs.10 and 11 we show the results of calculations of the differential 
cross-section of neutral current (NC) $e^{\pm}p$
scattering $d\sigma /dQ^2$ and $d\sigma /dy$ at high $Q^2$.
 The cross-section $d\sigma /dQ^2$ decreases by
six orders of magnitude between $Q^2=400$ GeV$^2$ and 4000 GeV$^2$. This
decrease is due to the photon propagator. The cross-section $d\sigma /dy$
is shown for different $Q^2$ regions. For $Q^2>400$ GeV$^2$ the bulk of the
cross-section is concentrated at small values of $y$. For $Q^2>10^4$ GeV$^2$
the differential cross-section is approximately constant with $y$. The 
predictions using the MRS99 PDF give a good 
%
description of measured cross-sections. NC scattering at high $Q^2$ is 
sensitive to the contribution due to the $Z^0$. According to Eqs.(4)-(9) 
the $Z^0$ contribution reduces approximately by 25\%(12\%) for $e^{+}p$
($e^{-}p$) cross-section at $Q^2>10^4$ GeV$^2$.
\section{Nuclear structure functions}  
\label{sect4}
The SFs measured for different nuclei $A$ are found to differ from
the SF measured on deuteron [24,25]. The modifications are usually observed 
as a deviation from unity of the ratio 
$r^{A/d}=F_2^{A}/F_2^d$, where $F_2^A$ and $F_2^d$ are the
SFs per nucleon measured in a nucleus and in deuteron, respectively. Neglecting
nuclear effects in the deuteron, $F_2^d$ can approximately stand for a isospin
averaged nucleon SF, $F_2^N=\left(F_2^p+F_2^n\right)/2$. Different nuclear
effects are observed in different region of $x$. 

(i) Shadowing at $x<$0.1. The ratio $r^{A/d}$ is smaller than unity. The 
experimental data cover the region $x>$10$^{-4}$ and  $r^{A/d}$ decreases
with decreasing $x$. Shadowing increases with nuclear mass $A$ and  
weakly depends on $Q^2$.

(ii) Anti-shadowing at $0.1<x<0.2$. The NMC data have established a small
(a few percent) but statistically significant excess over unity.
Within the accuracy of the data no significant $Q^2$-dependence of this
effect has been found.

(iii) EMC effect at $0.2<x<0.8$. The measured ratio $r^{A/d}$ decreases as 
 $x$ rises and has a minimum at $x=0.6$. The magnitude of this depletion 
grows slowly with nuclear mass number. The data imply that a strong 
$Q^2$-dependence of the $r^{A/d}$ is excluded in this region also.

(iv) Fermi motion. At $x>0.8$ the ratio $r^{A/d}$ rises above unity but 
experimental information is rather scarce.

Investigations of differences between the longitudinal-to-transverse 
cross-section ratio $R=\sigma_L/\sigma_T$ for
different nuclei showed that $R^{A_i}-R^{A_j}$ is compatible with zero. This 
implies that nuclear effects influence both SFs $F_1$ and $F_2$ in a similar 
way. 

At the moment, there is no unique theoretical description of these effects;
it is believed that different mechanisms are responsible for them in
different kinematical regions. For example, the EMC effect indicates that the 
averaged momentum carried by valence quarks in nuclei is reduced, relative 
to free nucleon. It has been shown in [26] that the pattern of the
function $r^{A/d}(x)$ has a universal shape in the range $10^{-3}<x<0.96$
and in the range of nuclei mass $A\ge$4. Namely, the ratio $F_2^A(x)/F_2^d(x)$
could be well approximated with phenomenological functions in different regions
 of $x$. At $x>0.3$
\begin{equation}
r^{A/d}(x)=1-m_b(A)a_{osc}(x),     
\end{equation}
where the $A$ dependence of $m_b$ can be approximated as
\begin{equation}
m_b(A)=M_b\left(1-N_s(A)/A\right)\mbox{~~~~~~and~~~~~~}M_b=0.437     
\end{equation}
for $A\ne$4. The number of nucleons $N_s(A)$ at the nucleare surface is 
given by the Woods-Saxon potential
\begin{equation}
N_s(A)=4\pi\rho_0\int\limits_{r_0(A)}^{\infty}\frac{r^2dr}{1+exp\left\{
(r-r_0(A))/a\right\}}                                    
\end{equation}
with values of parameters $\rho_0$=0.17 fm$^{-3}$, $a$=0.54 fm and
\begin{equation}
r_0(A)=1.12A^{1/3}-0.86A^{-1/3}.                     
\end{equation}
The function $a_{osc}(x)$ is
\begin{equation}
a_{osc}(x)=(1-\lambda x)\left\{\left(\frac{1}{u}-\frac{1}{c}\right)-\mu\left(
\frac{1}{u^2}-\frac{1}{c^2}\right)\right\},                      
\end{equation}
where $u=1-x$, $c=1-x_2$, $x_2=$0.278, $\lambda$=0.5 and $\mu=m_{\pi}/M$
 ($m_{\pi}$ is pion mass).
At 10$^{-3}\le x\le$0.3 the function is given by 
\begin{equation}
r^{A/d}(x)=x^{m_1}(1+m_2)(1-m_3x),                                
\end{equation}
with
\begin{equation}
m_i=M_i\left(1-N_s(A)/A\right),                                
\end{equation}
where $M_1$=0.129, $M_2$=0.456 and $M_3$=0.553. We used Eq.(49) for 
calculation  of $r^{A/d}$ up to $x_0<10^{-3}$. 

The value of $x_0$ as a function of A was obtained in the following way.
The experimental data [26], show that in the region $5\cdot 10^{-3}<x<0.1$ 
the ratio $r^{A/d}$ decreases with $x$. Generally, small $x$ corresponds to
small $Q^2$, therefore the approach of real photon interaction can be used.   
So, for $x\to 0$, 
$r^{A/d}\to \eta^{A}$=$\sigma_{\gamma A}/A\sigma_{\gamma N}$, where 
$\sigma_{\gamma A}$ is photon-nuclear cross-section and $\sigma_{\gamma N}$ 
is the photon-nucleon cross section averaged over proton and neutron. 
The expression for the function $\eta^A$  has been obtained in Ref.[4], 
using the optical nuclear model 
\begin{equation}
\eta^A=0.75G(z)+0.25,                                
\end{equation}
where the function $G(z)$ is
\begin{equation}
G(z)=\frac{3}{z^2}\left[\frac{z^2}{2}-1+e^{-z}(1+z)\right],    
\end{equation}
and $z=0.00282A^{1/3}\sigma_{\gamma N}(\nu)$.
Using Eq.(35) with the values of parameters obtained in this work, we can write
 the averaged photon-nucleon cross-section as follows 
\begin{equation}
\sigma_{\gamma N}=\frac12(\sigma_{\gamma p}+\sigma_{\gamma n})=
112.2\left(0.609\nu^{0.0988}+1.037\nu^{-0.5944}\right).           
\end{equation}
In the range $x\ll1$, Eq.(49) reduces to
\begin{equation}
r^{A/d}(x)=x^{m_1}(1+m_2).                                
\end{equation}
Then, from the asymptotic condition
\begin{equation}
r^{A/d}(x_0)=0.75G(z)+0.25                                
\end{equation}
we obtain the expression for $x_0$ 
\begin{equation}
x_0=\left[\frac{1}{1+m_2}\left(0.75G(z)+0.25\right)\right]^{1/m_1}.   
\end{equation}
At $x<x_0$ we assumed that the function $r^{A/d}$ is constant and 
\begin{equation}
r^{A/d}(x)=r^{A/d}(x_0).                            
\end{equation}
The results of approximation of the ratio $r^{A/d}$ are presented in Fig.12
 as a function of $x$ for different nuclear targets and are in a good  
agreement with  data.  

Taking into account the nuclear effects, the nuclear SFs $F_i(x,Q^2)$ and 
total photon-nuclear cross section can be written as:
\begin{equation}
F_i^A(x,Q^2)=Ar^{A/d}(x,A)F_i^N(x,Q^2)               
\end{equation}
and
\begin{equation}
\sigma_{\gamma A}(\nu)=A\sigma_{\gamma N}(\nu)[0.75G(z)+0.25].  
\end{equation}
The calculated cross sections $\sigma_{\gamma A}(\nu)$ are shown in Fig.13 
as functions of the real photon energy for nuclei 
 C, Cu and Pb.
\\
\section{Muon inelastic scattering in standard rock} 
\label{sect5}
We have calculated the main characteristics of inelastic muon scattering in
 standard rock ( $A=22$, $Z=11$ and $\rho=2.65$ g/cm$^3$ ).
The spectra of muon energy loss ($N_{Av}$ is Avogadro's number) in single
interaction,
\begin{equation}
\frac{N_{Av}}{A}\nu\frac{d\sigma_{\mu A}}{d\nu}=\frac{N_{Av}}{A}\nu\int
\limits_{Q^2_{min}}^{Q^2_{max}}\frac{d\sigma_{\mu A}^2}{d\nu dQ^2}dQ^2,  
\end{equation}
are shown in Fig.14 as functions of inelasticity $y$ for different muon 
energies. The energy dependence of the total cross section,
\begin{equation}
\sigma_{\mu A}(E)=\int\limits_{\nu_{min}}^
{\nu_{max}}\frac{d\sigma_{\mu A}}{d\nu}d\nu,              
\end{equation}
and muon energy loss,
\begin{equation}
b_n(E)=\frac{N_{Av}}{A}\int\limits_{\nu_{min}}^
{\nu_{max}}\nu\frac{d\sigma_{\mu A}}{d\nu}d\nu,              
\end{equation}
are shown in Fig.15. 
The allowed kinematical region for the variables $\nu$ and $Q^2$ is
determined by the following equations:
\begin{equation}
Q^2=2\left(EE' - \mid{k}\mid\mid{k'}\mid \cos{\theta}\right) - 2m^2  
\end{equation}
at $\cos{\theta}=\pm1$, and
\begin{equation}
Q^2=2M\nu+M^2-W^2.                                              
\end{equation}
The results from Ref.[4] are given in these figures for 
comparison also. It is necessary to note, that the 
cross-section and muon energy loss [4] have been calculated taking into 
account the shadowing effect only. 
The cross-section of inelastic muon scattering obtained in the present work 
is larger by factor 1.2 and the muon energy losses $b_n(E)$ is larger too 
by $\approx$8\% at $E=$10$^3$ GeV and by $\approx$30\% at $E=$10$^6$ GeV. 
As a result, the total energy losses (the sum of bremsstrahlung, 
pair-production and inelastic muon scattering) increases by $\approx$1\% 
at $E=$10$^3$ GeV and by $\approx$4\% at $E=$10$^6$ GeV. These differences are
 mainly due to contributions of small $x$ and small $Q^2$ where the modern 
SFs are large than SFs used by Bezrukov and Bugaev.

The probabilities $P(\ge\theta,\ge v)$ of muon scattering in single
interaction at the angles larger than $\theta$ with outgoing muon energy 
$E'\ge vE$ are shown in Fig.16 as a function of $\theta$ for different values 
of $v$ and primary muon energies. The results are given for $\mu^{-}$ and 
$\mu^{+}$ scattering. The main peculiarities of the inelastic muon scattering 
are: 

(i) At fixed values of $\theta$ and $E'$ the probability decreases 
very quickly with the initial muon energy $E$. For example, for 
$\theta\ge 2^\circ$ and $E'$=10 GeV, $P$=6.3$\cdot$10$^{-4}$ at 
$E$=10$^2$GeV and $P=$3.8$\cdot$10$^{-6}$ at $E$=10$^{3}$GeV.

(ii) At fixed values of $\theta$ and $E$ the probability increases with 
decreasing outgoing muon energy $E'$. For $\theta\ge$2$^\circ$ and 
$E$=10$^3$GeV, $P$=7.8$\cdot$10$^{-7}$ at $E'\ge$10$^2$GeV and 
$P=$3.8$\cdot$10$^{-6}$ at $E'\ge$10GeV.  

(iii) At fixed values of $E$ and $E'$ the mean values of $x$ and $Q^2$
($\langle x\rangle$ and $\langle Q^2\rangle$) increase with $\theta$. For muon 
energies $E$=10$^2$ GeV and $E' \ge$0.1$E$ the values of $\langle x\rangle$ 
and $\langle Q^2\rangle$ increase from $\langle x\rangle$=0.12
 (anti-shadowing region) and $\langle Q^2\rangle$=0.75 GeV$^2$ at 
$\theta$=0.25$^{\circ}$ up to $\langle x\rangle$=0.25 (EMC-region) and 
$\langle Q^2\rangle$=34 GeV$^2$ at $\theta$=6$^{\circ}$. For energies 
$E$=10$^3$ GeV and $E' \ge$0.1$E$: $\langle x\rangle$=0.09 (shadowing region),
$<Q^2>$=28 GeV$^2$ at $\theta$=0.25$^{\circ}$ and $\langle x\rangle$=0.46
(anti-shadowing region) and  $\langle Q^2\rangle$=925 GeV$^2$ at 
$\theta$=6$^{\circ}$. Thus, the probability of scattering at large angles
is suppressed by the EMC-effect. 
\section{Conclusions} 
\label{sect6}
In this work we studied inelastic muon scattering off nuclei.

1. It is shown that the combination of the SF $F_2$, predicted by the CKMT 
model at low and moderate values of $Q^2$, and the MRS99 PDF at high $Q^2$ 
gives a good description of the data over the complete measured region from 
photo-production to DIS. In particular, the CKMT model well describes the 
rise of the $F_2^p$ at low $x$ and $Q^2$ with the slope $\Delta_0=0.0988$. 
Furthermore, in the framework of this model the expression for the neutron 
SF $F_2^n$ can be obtained. The result is in a good agreement with the 
$F_p^2/F_n^2$ and $F_p^2-F_n^2$ data.\

2. The MRS99 PDF well describe the differential cross-sections 
$d\sigma/dQ^2$ and $d\sigma/dy$, calculated taking into account not only
electromagnetic current contribution but also contributions of neutral 
current and $\gamma-Z$ interference. The $\gamma-Z$ interference contribution
is clearly seen at high $Q^2>10^3$ GeV$^2$.

3. The nuclear effects modify the nucleon SFs in the entire measured region 
of $x$ and $Q^2$. The modification depends very slowly on $Q^2$ and increases 
with $A$.

4. The obtained SFs have been used for calculations of the muon-nucleus 
scattering cross-section, muon energy losses and muon angular distributions
in inelastic interaction taking into account nuclear effects and contributions
of neutral current and $\gamma-Z$ interference. As a result, the total 
cross-section and energy losses increase with muon energy faster then 
predicted in Ref.[4]. The scattering of high energy muons ($E>10^3$) GeV at 
large angles is suppressed by EMC-effect and $\gamma-Z$ interference.

\section{Acknowledgements} 
We are grateful to A.B. Kaidalov and F.V. Tkachov for a helpful discussions. 
This work was supported by Russian Foundation for Basic Research grant 
number 99-02-18374.

\newpage
\begin{center}
\mbox{\epsfig{file=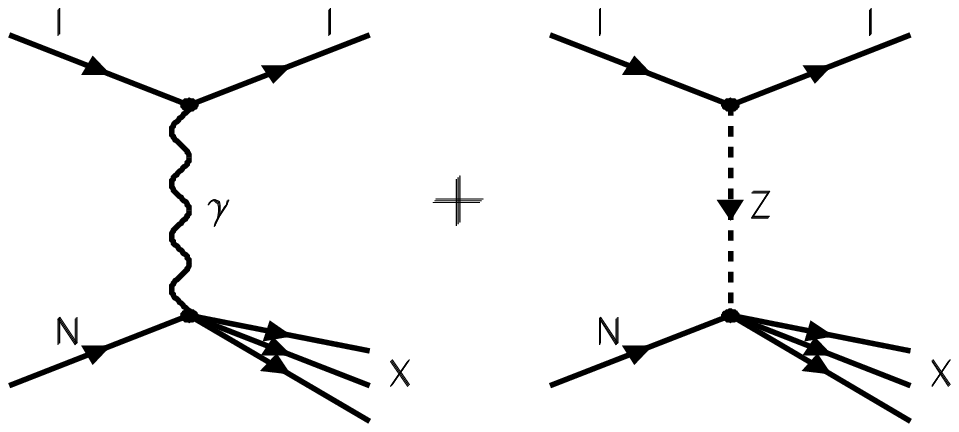,height=10cm,width=14cm}}
\vspace{2mm}
\noindent
\small
\end{center}
{\sf Fig.~1:}~Schematic diagrams for neutral current charged lepton 
scattering off nucleon.
\newpage
\begin{center}
\mbox{\epsfig{file=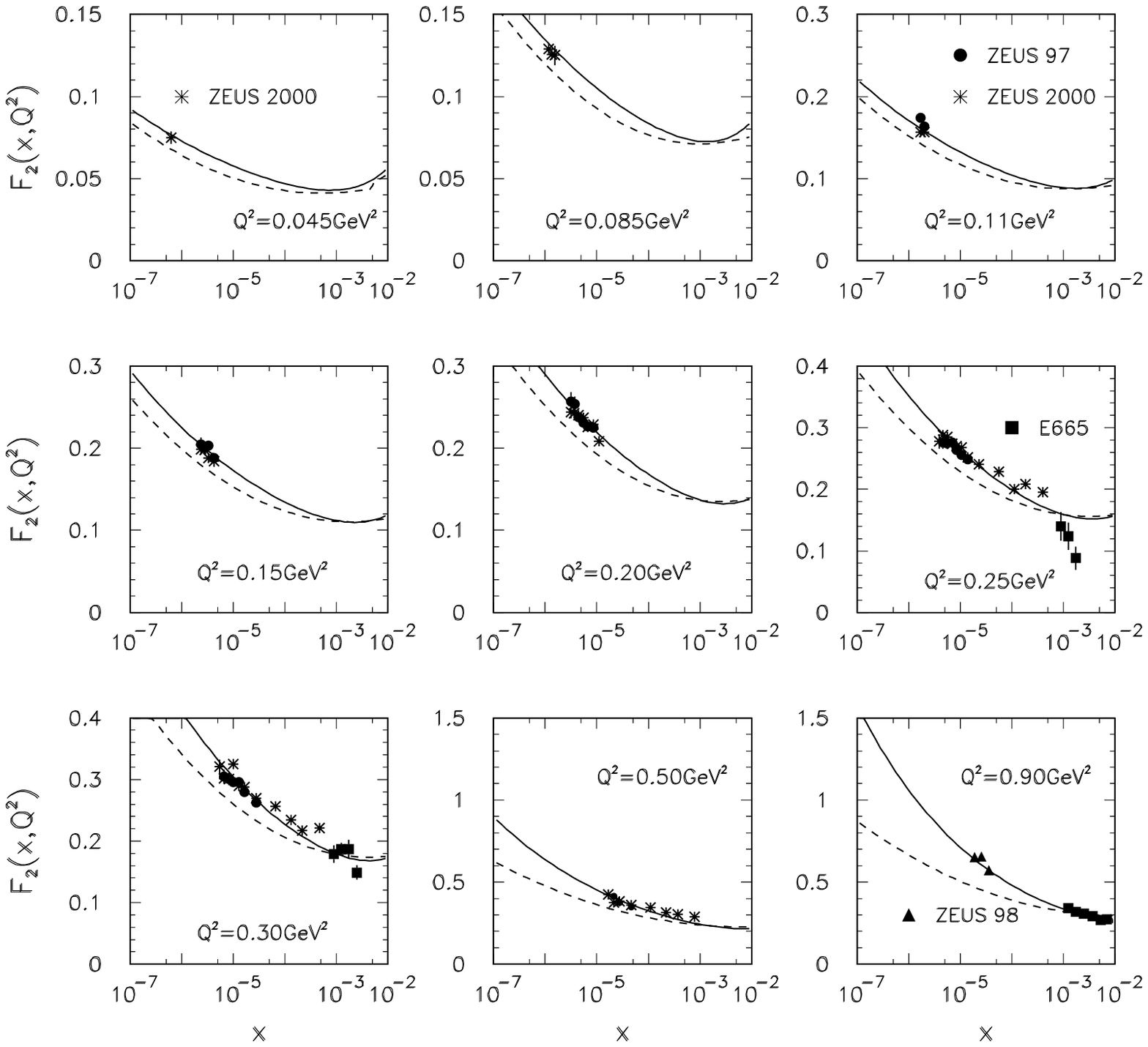,height=15cm,width=16cm}}
\vspace{2mm}
\noindent
\small
\end{center}
{\sf Fig.~2a:}~The proton structure function $F_2^p$ at low $Q^2$ as a 
function of $x$. Solid lines are our calculations using the CKMT model and 
dashed lines calculations of Bezrukov and Bugaev [4]. Points are results of 
experiments ZEUS and E665 [5,17].    
\newpage
\begin{center}
\mbox{\epsfig{file=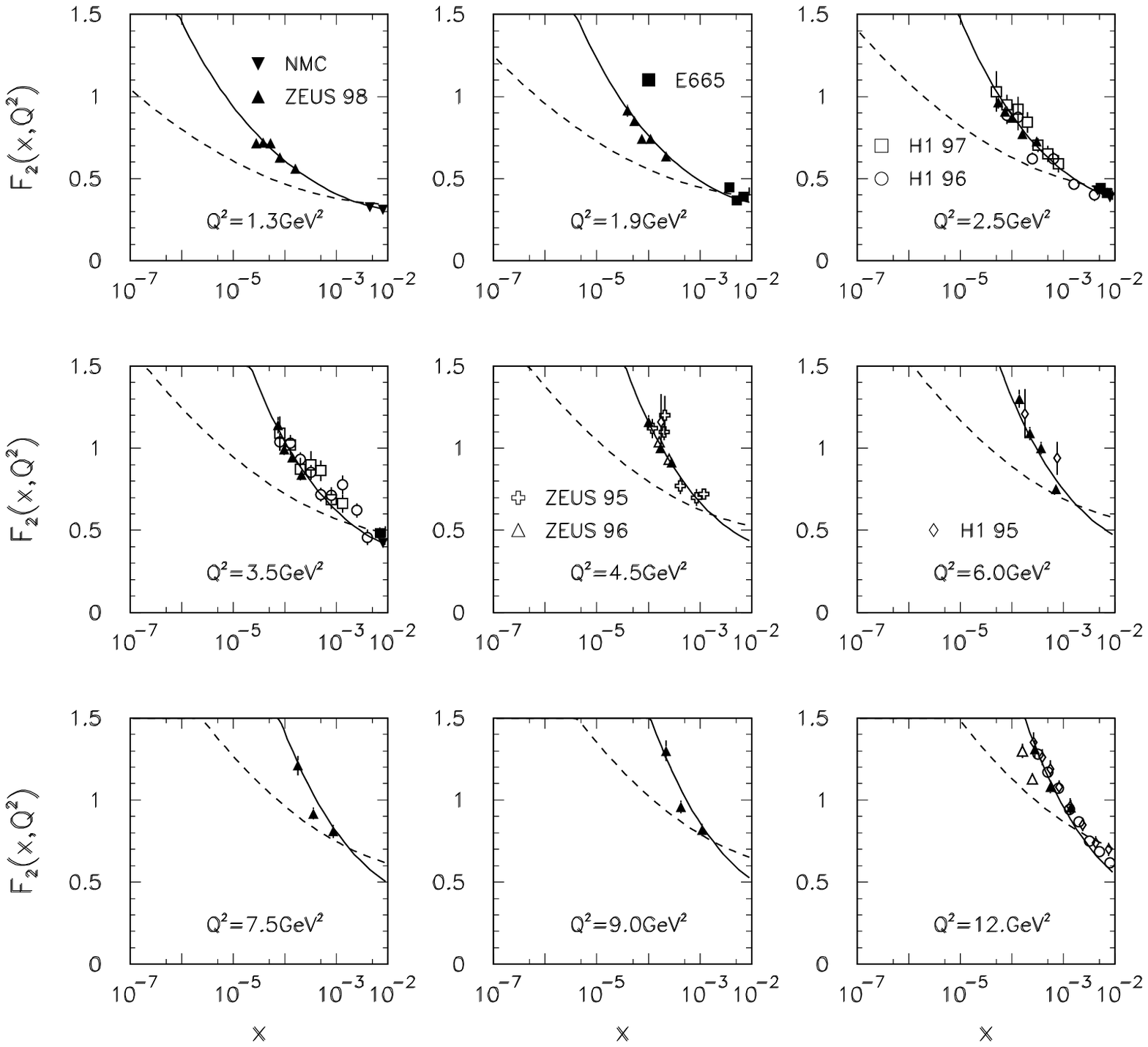,height=15cm,width=16cm}}
\vspace{2mm}
\noindent
\small
\end{center}
{\sf Fig.~2b:}~The same as Fig.2a but at moderate $Q^2$.
\newpage
\begin{center}
\mbox{\epsfig{file=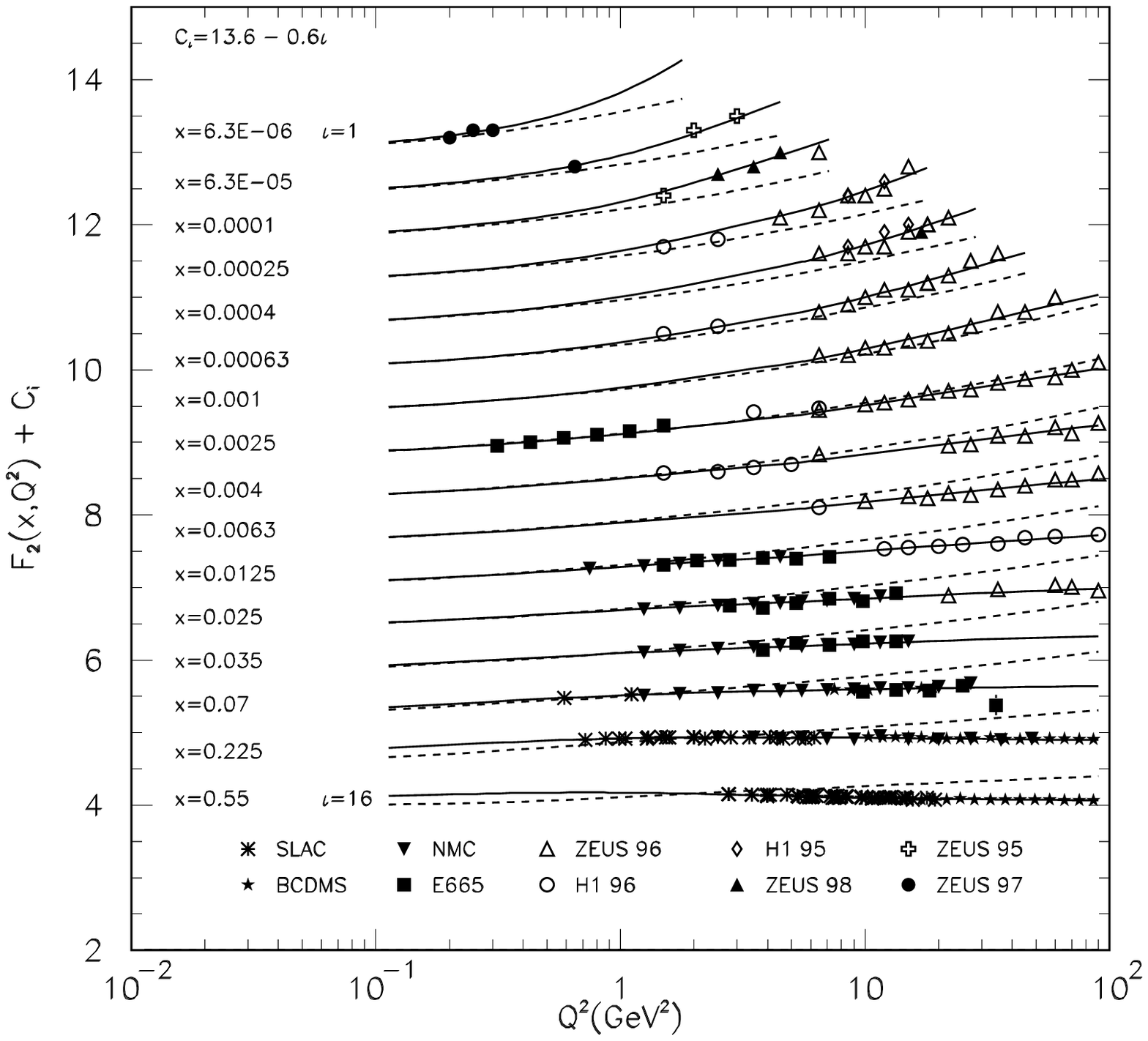,height=14cm,width=16cm}}
\vspace{2mm}
\noindent
\small
\end{center}
{\sf Fig.~3:}~The proton structure function $F_2^p$ as a function of $Q^2$ at 
fixed values of $x$. The solid curves have been obtained by using the 
CKMT+MRS99 model and dashed curves are result from Ref.[4]. Data set is due to
 ZEUS, H1, E665, NMC, and SLAC experiments [5]. For clarity an amount 
$C_i$=13.6-0.6$i$ is added to $F_2^p$ where $i$=1(16) for the lowest (highest)
 $x$ value. 
\newpage
\begin{center}
\mbox{\epsfig{file=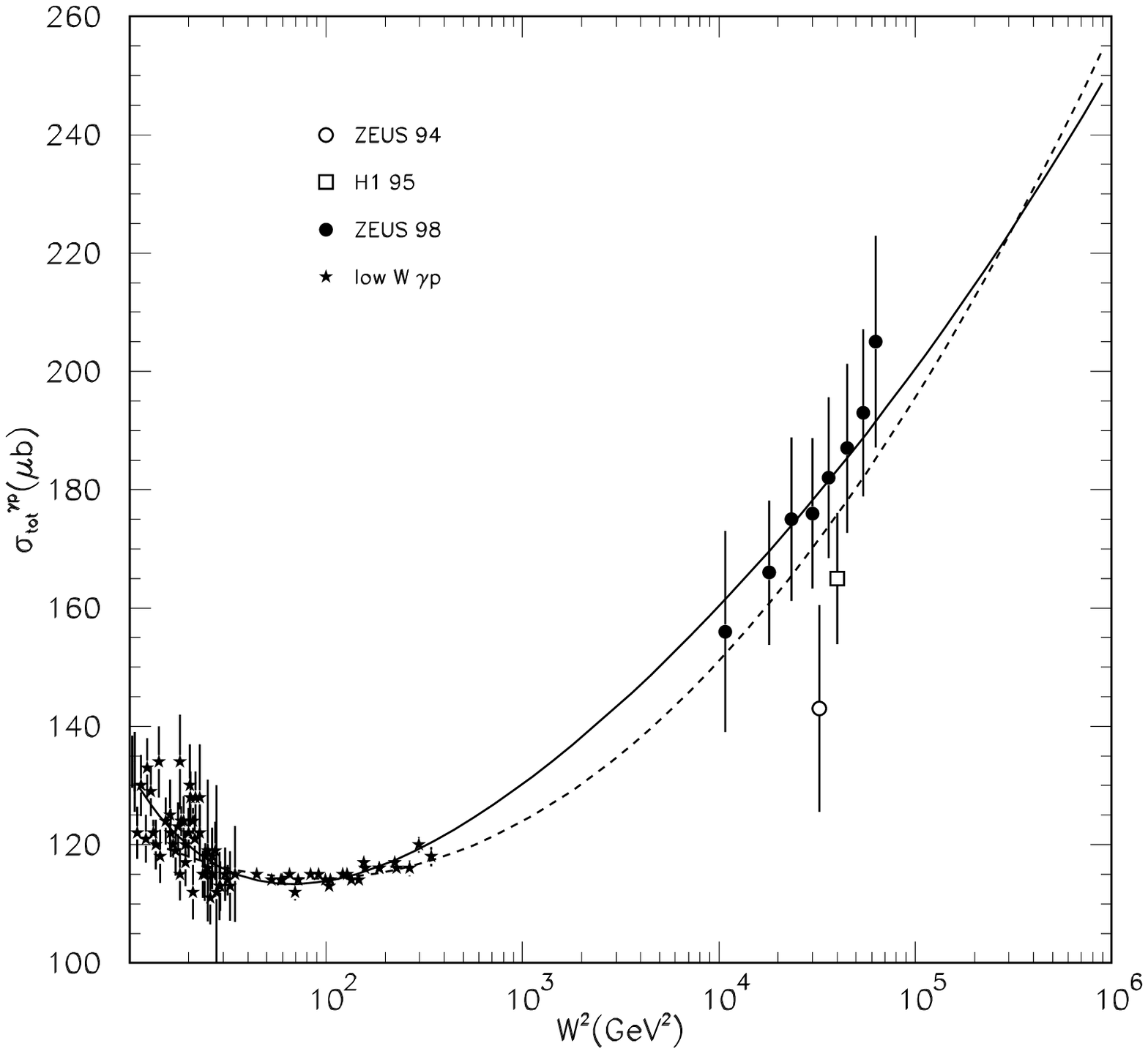,height=15cm,width=16cm}}
\vspace{2mm}
\noindent
\small
\end{center}
{\sf Fig.~4:}~Total cross-section $\sigma^{\gamma p}_{tot}$ as a function of 
$W^2$. Solid line is our calculations using combined (CKMT+MRS99) model and 
dashed one is from work [4]. Experimental data are from [15] at low energies 
and from ZEUS and H1 [16] at higher energies.
\newpage
\begin{center}
\mbox{\epsfig{file=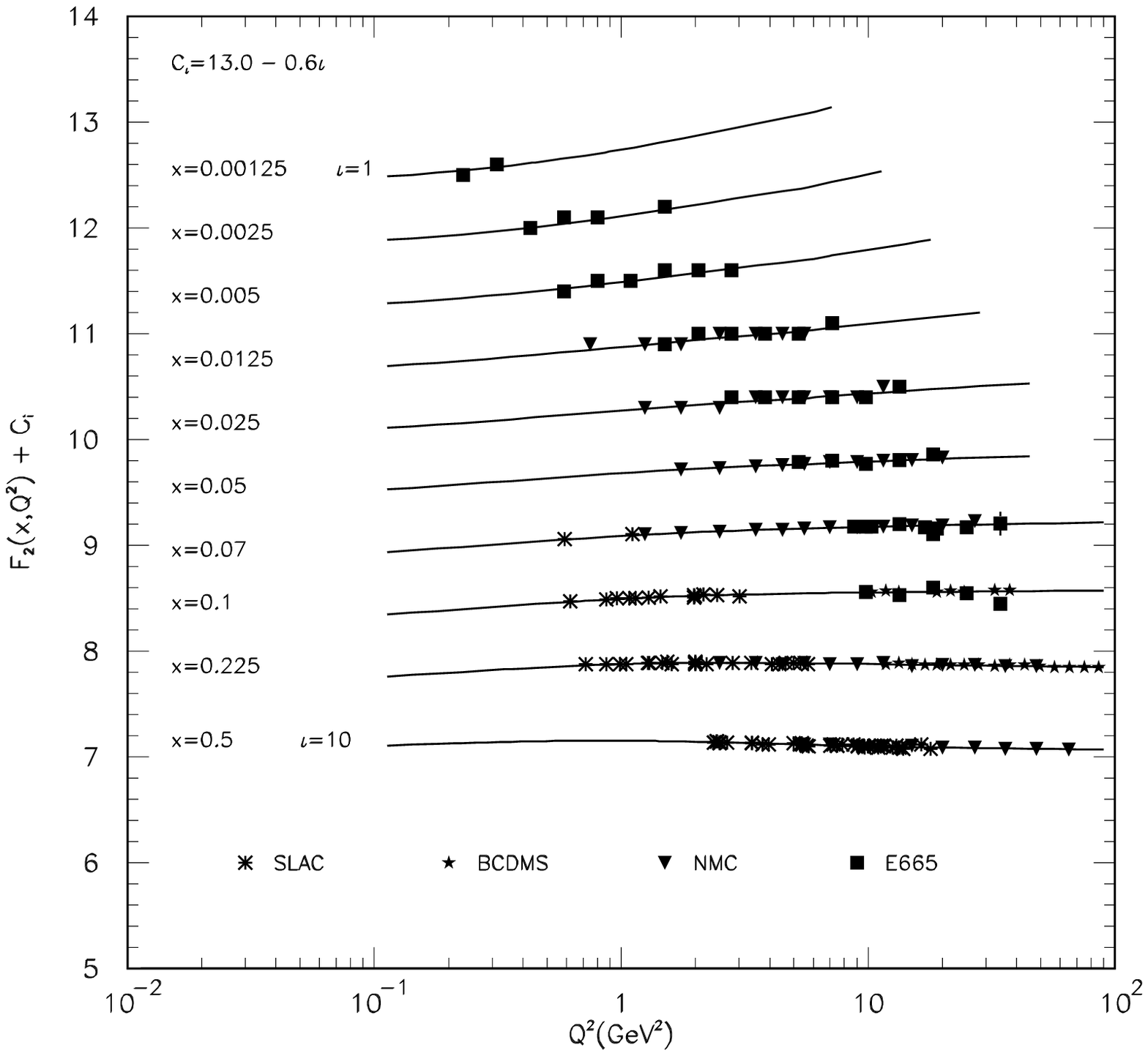,height=15cm,width=16cm}}     
\vspace{2mm}
\noindent
\small
\end{center}
{\sf Fig.~5:}~The deutron structure function $F_2^d$ as a function of $Q^2$ at 
fixed values of $x$. Data set is due to BCDMS, E665, NMC, and SLAC experiments [5]. For clarity an amount $C_i$=13.6-0.6$i$ is added to $F_2^d$ where 
$i$=1(10) for the lowest (highest) $x$ value.           
\newpage
\begin{center}
\mbox{\epsfig{file=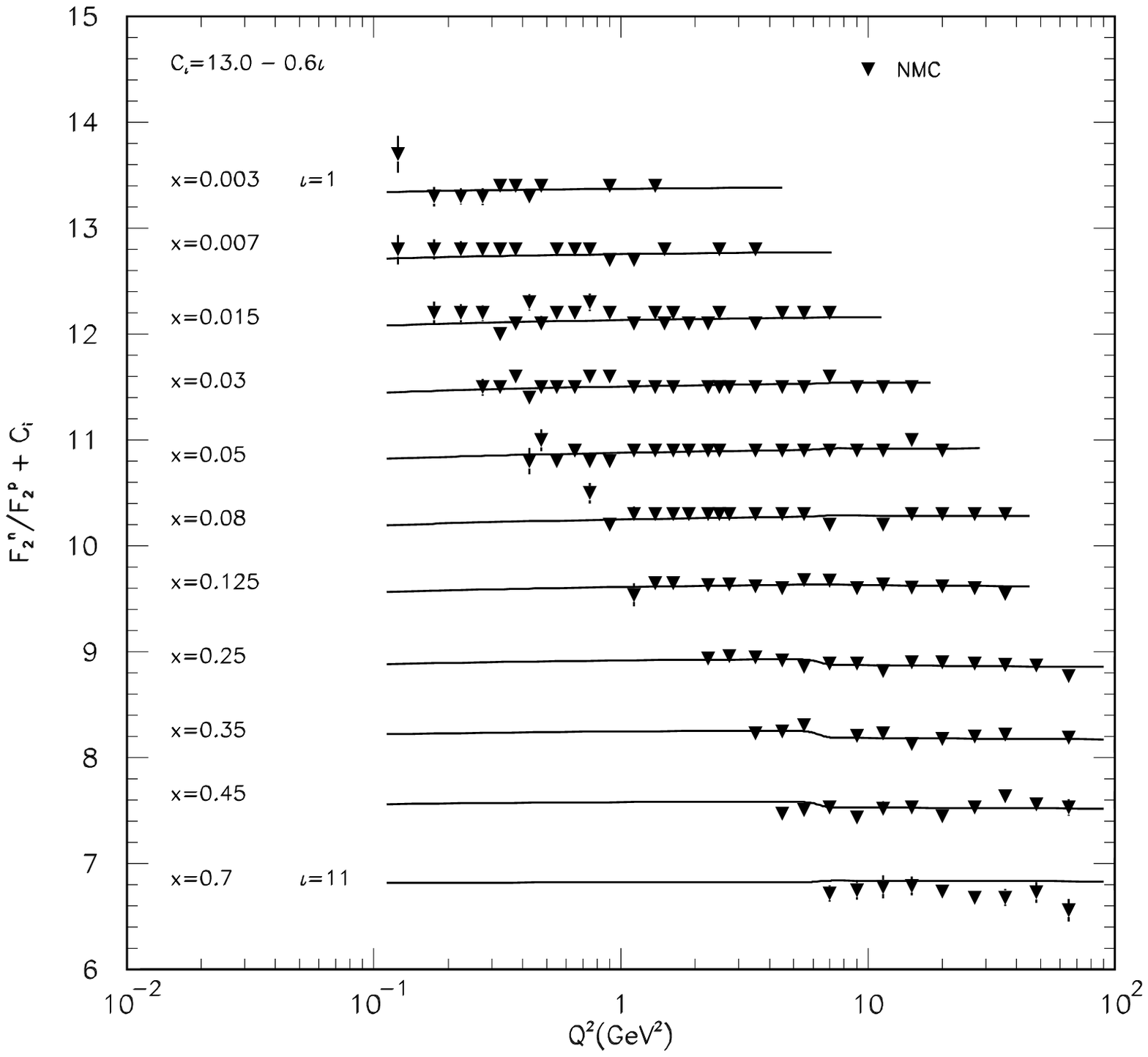,height=15cm,width=16cm}}
\vspace{2mm}
\noindent
\small
\end{center}
{\sf Fig.~6:}~The ratio of structure functions $F_2^n/F_2^p$ as a function of
$Q^2$ at fixed values of $x$, as compared to NMC data [5]. For clarity an 
amount $C_i=13.6-0.6i$ is added to $F_2^n/F_2^p$ where $i$=1(11) for the 
lowest (highest) $x$ value.           
\newpage
\begin{center}
\mbox{\epsfig{file=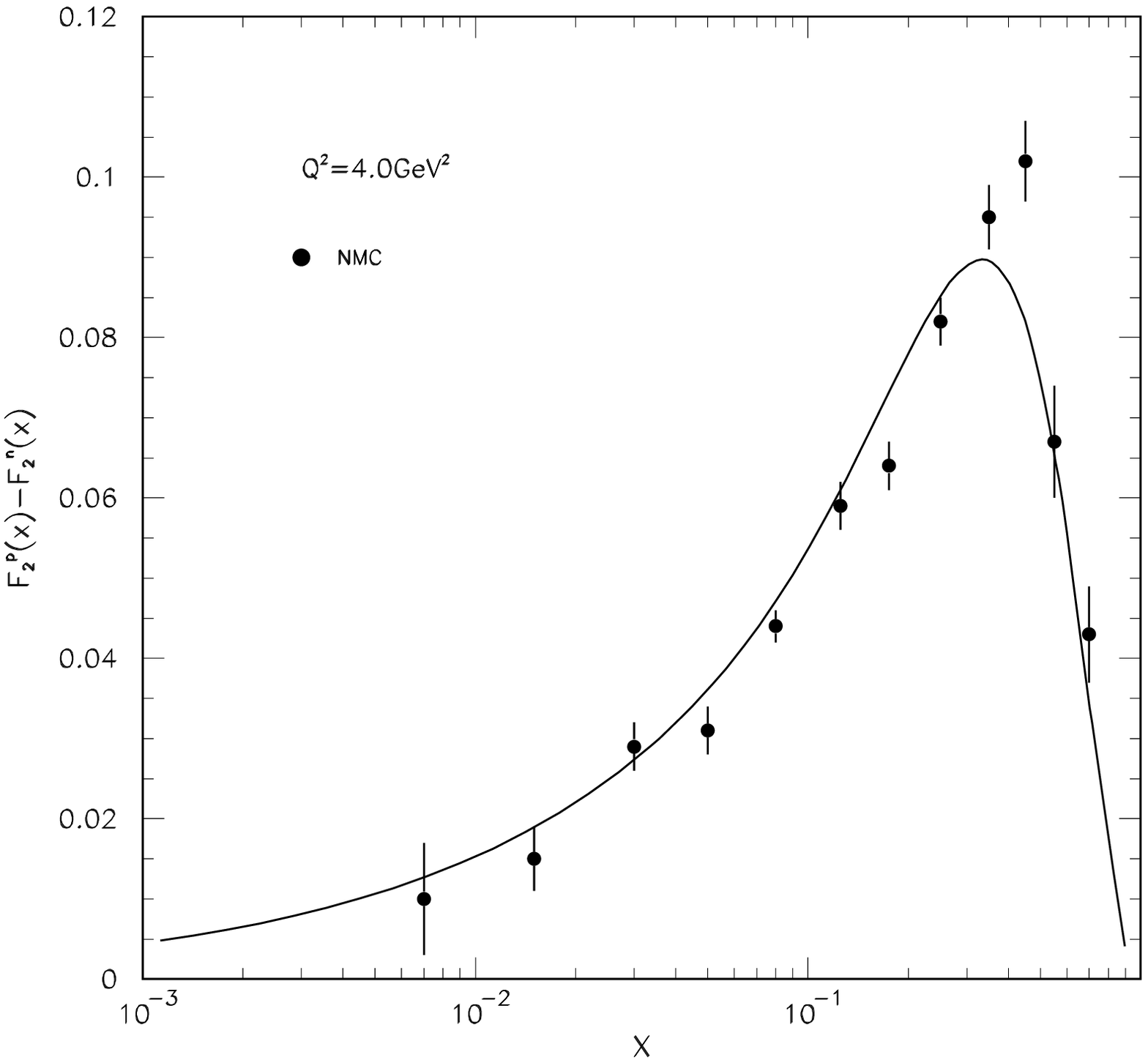,height=15cm,width=16cm}}
\vspace{2mm}
\noindent
\small
\end{center}
{\sf Fig.~7:}~The difference $F_2^p-F_2^n$ at $Q^2$=4GeV$^2$ as a function 
of $x$, as compared to NMC data [5].   
\newpage
\begin{center}
\mbox{\epsfig{file=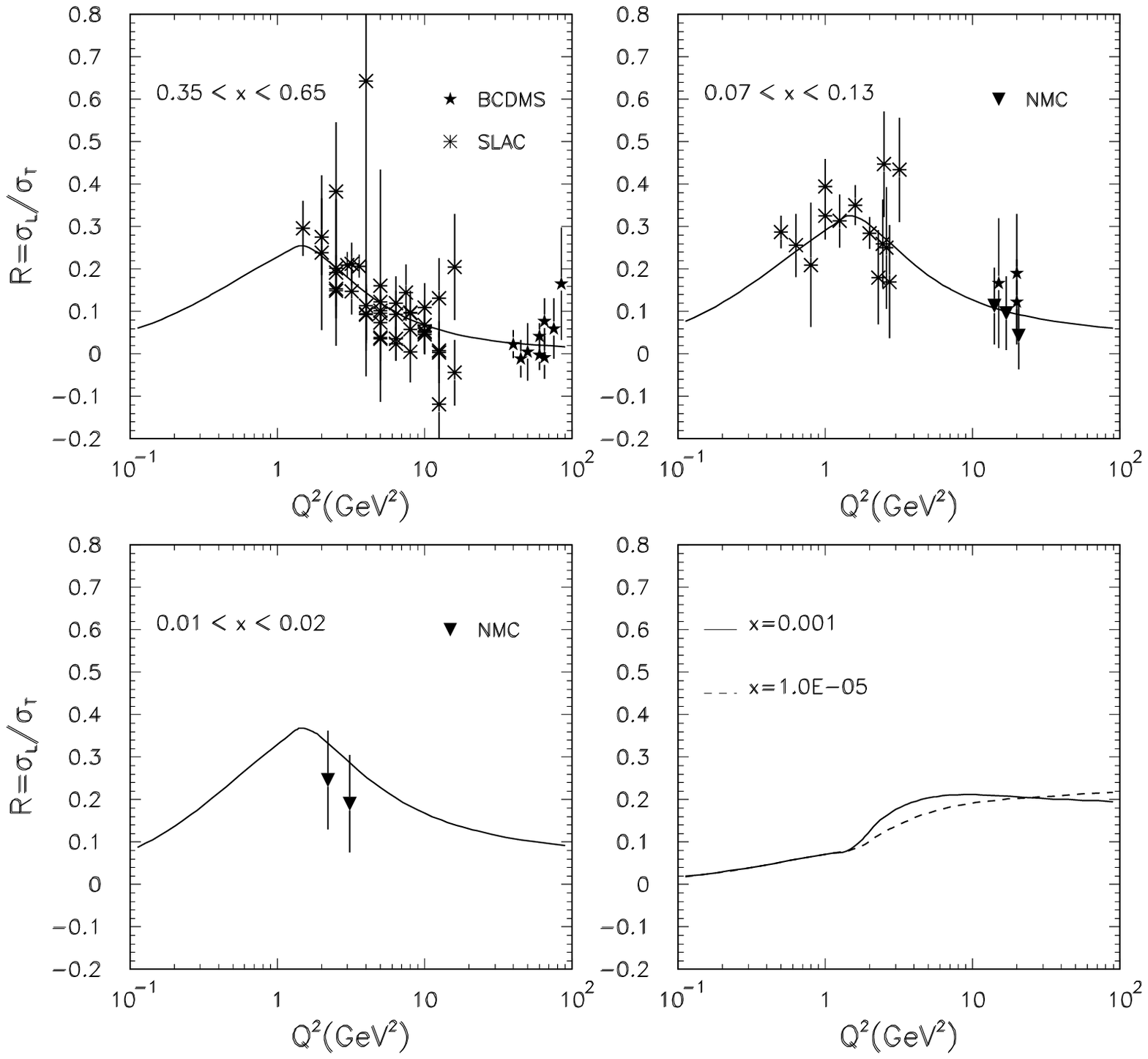,height=15cm,width=16cm}}
\vspace{2mm}
\noindent
\small
\end{center}
{\sf Fig.~8:}~The ratio $R(x,Q^2)$ as a function of $Q^2$ at fixed $x$. Data 
are from Refs.[5] and [20].              
\newpage
\begin{center}
\mbox{\epsfig{file=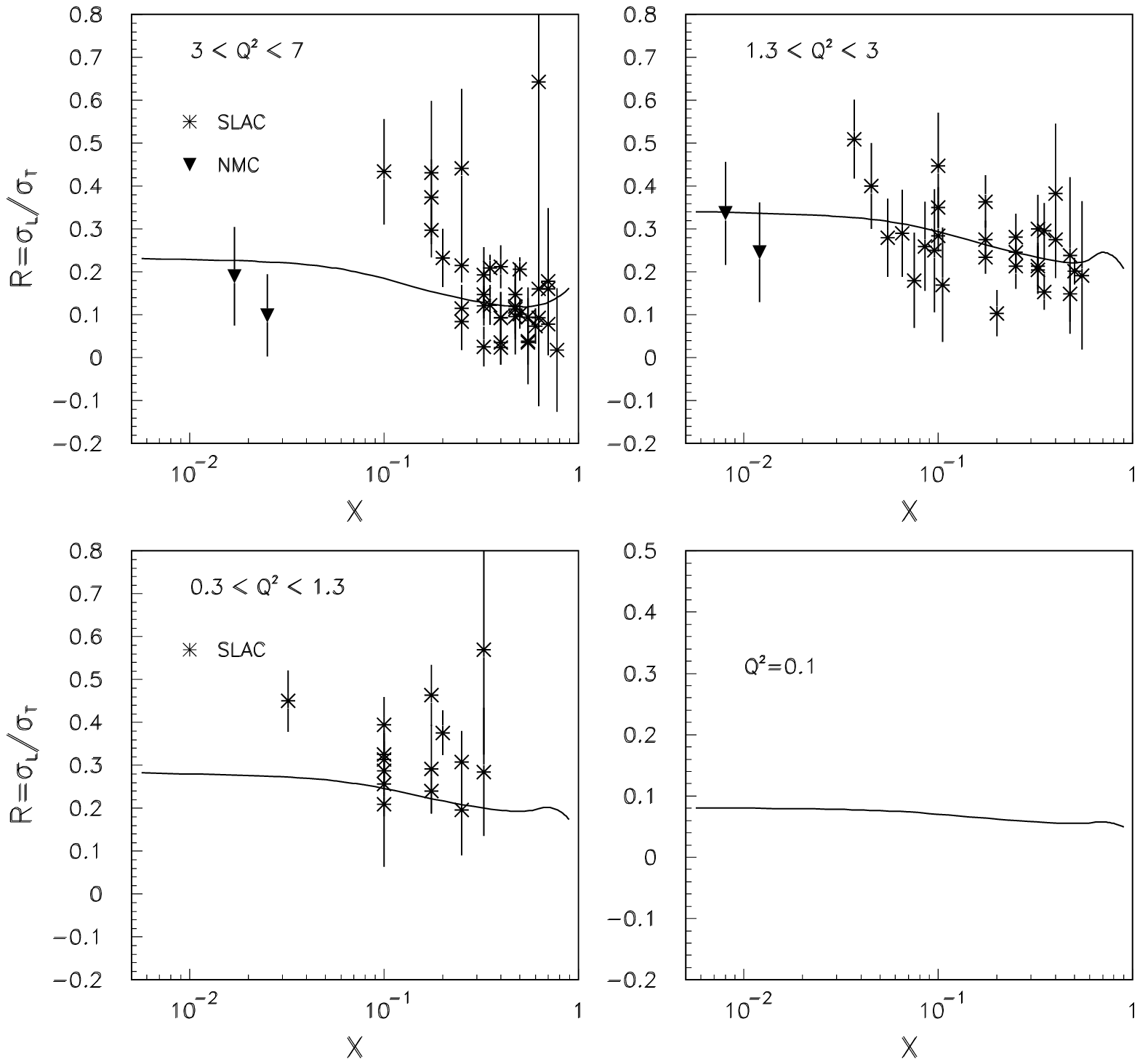,height=15cm,width=16cm}}
\vspace{2mm}
\noindent
\small
\end{center}
{\sf Fig.~9:}~The ratio $R(x,Q^2)$ as a function of $x$ at fixed $Q^2$. Data 
are from Refs.[5] and [20].              
\newpage
\begin{center}
\mbox{\epsfig{file=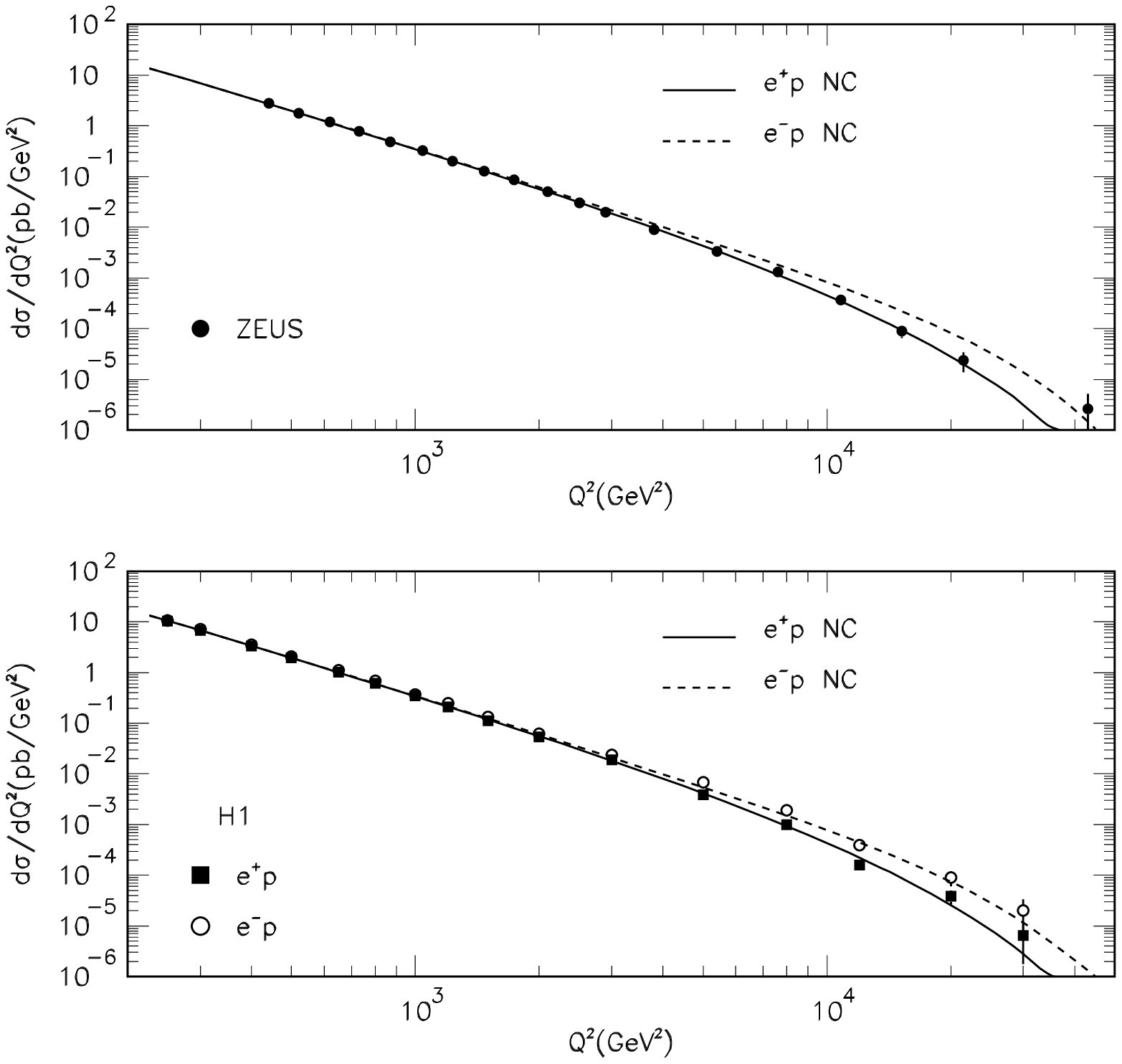,height=15cm,width=16cm}}
\vspace{2mm}
\noindent
\small
\end{center}
{\sf Fig.~10:}~The differential cross-section $d\sigma /dQ^2$ of neutral 
current proton scattering off electron (positron) as a function of $Q^2$.
Points are experimental results of ZEUS [22] and H1 [23] experiments.  
\newpage
\begin{center}
\mbox{\epsfig{file=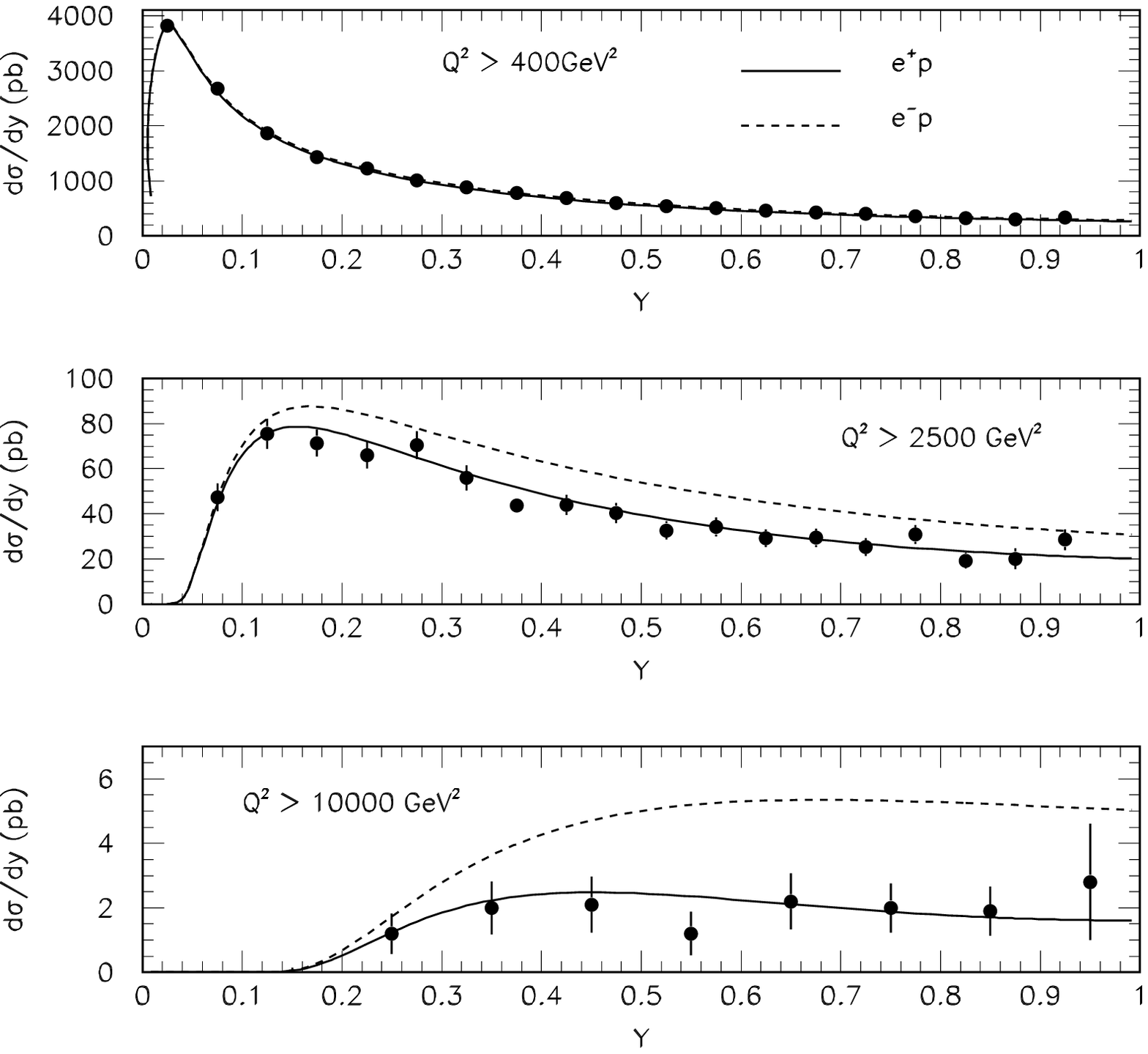,height=15cm,width=16cm}}
\vspace{2mm}
\noindent
\small
\end{center}
{\sf Fig.~11:}~The differential cross-section $d\sigma /dy$ of neutral
current proton scattering off electron (positron) as a function of $y$.
Experimental data are due to  ZEUS [22] experiment.  
\newpage
\begin{center}
\mbox{\epsfig{file=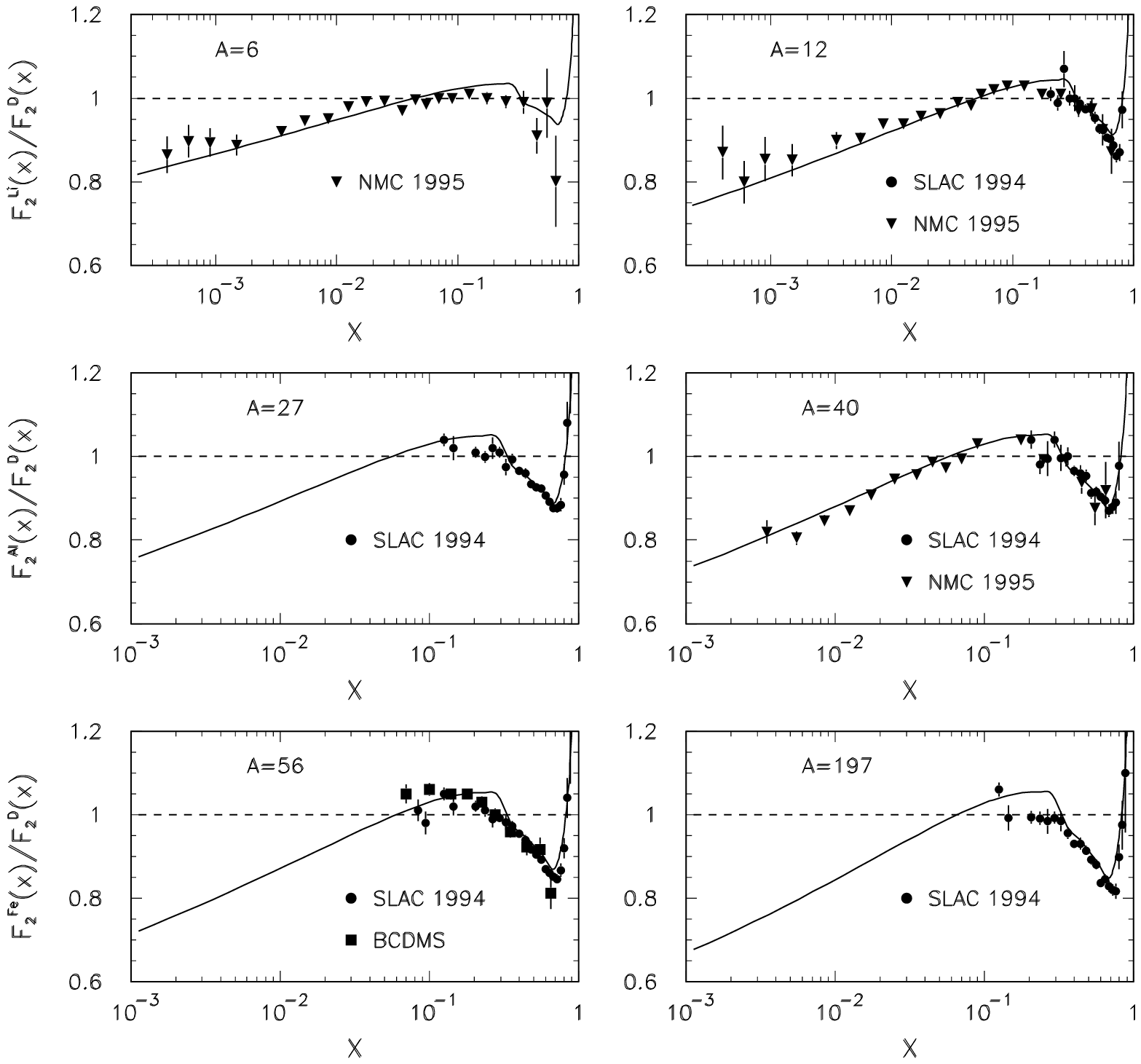,height=16cm,width=16cm}}
\vspace{2mm}
\noindent
\small
\end{center}
{\sf Fig.~12:}~The comparison of ratio $F_2^{A}/F_2^d$, measured by SLAC, 
BCDMS and NMC groups and approximations given in work [26].  
\newpage
\begin{center}
\mbox{\epsfig{file=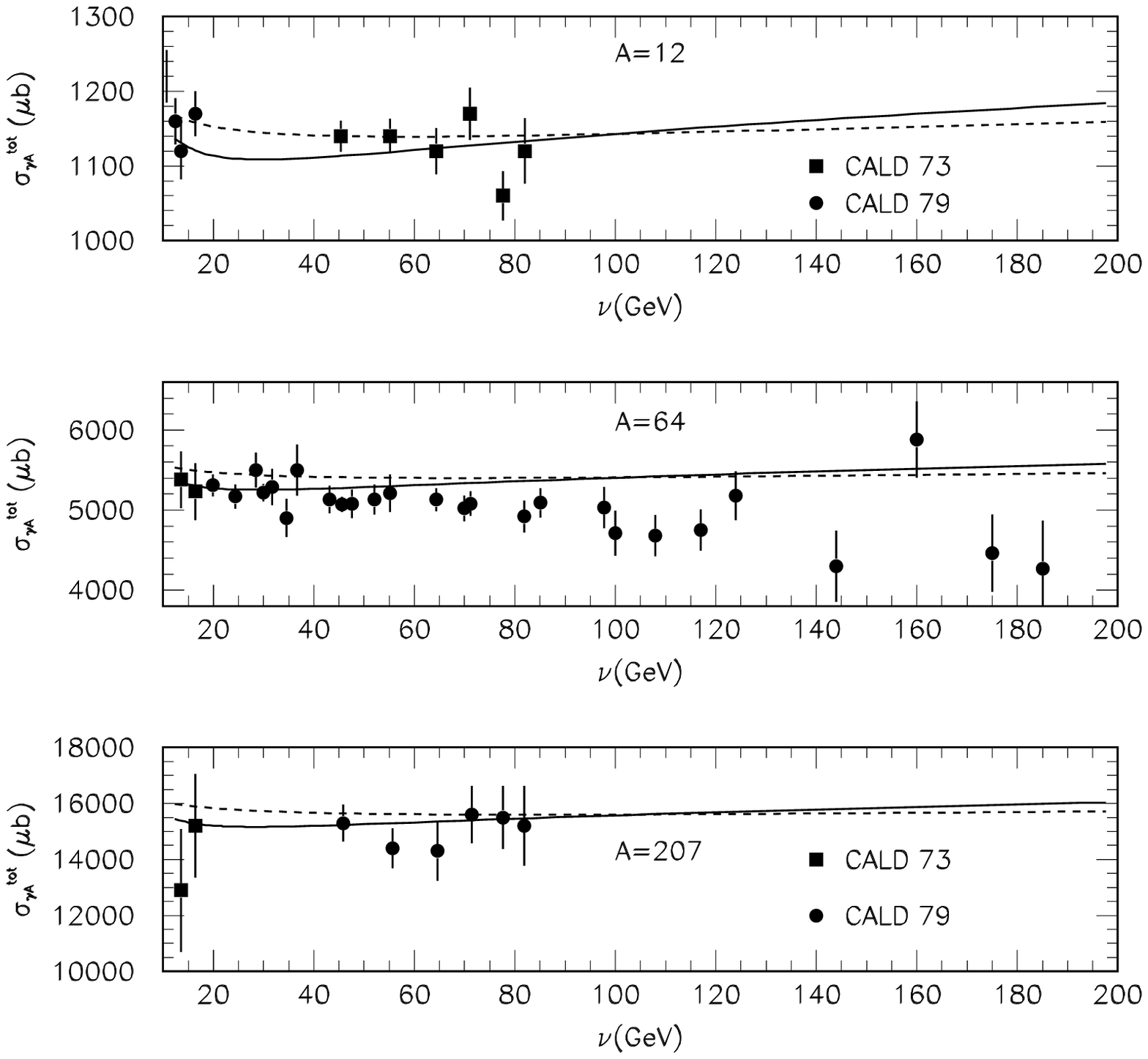,height=15cm,width=16cm}}
\vspace{2mm}
\noindent
\small
\end{center}
{\sf Fig.~13:}~ Total cross-section $\sigma_{\gamma A}(\nu)$ for C, Cu, and Pb
 as a function of real photon energy. Results of our calculations (solid 
curves) and calculations by Bezrukov and Bugaeev [4] (dashed curves) as 
compered to experimental data [27].
\newpage
\begin{center}
\mbox{\epsfig{file=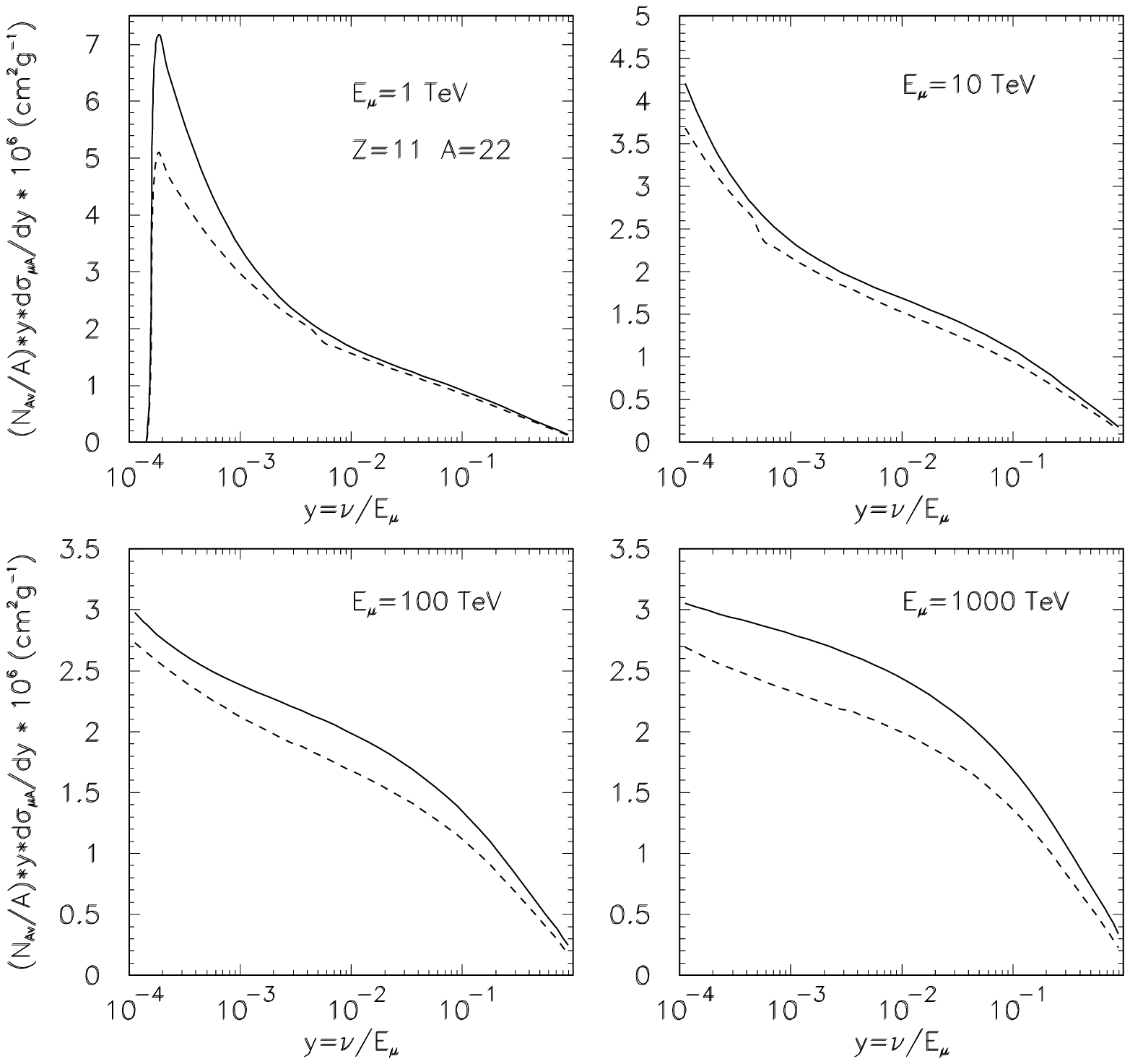,height=15cm,width=16cm}}
\vspace{2mm}
\noindent
\small
\end{center}
{\sf Fig.~14:}~The spectra of muon energy loss due to  
muon inelastic scattering in standard rock as a function of  $y$ for fixed 
 muon energies, as compered to calculations by Bezrukov and Bugaev 
[4]. ($y$ is the fraction of $E_{\mu}$ lost by muon in single interaction.) 
\newpage
\begin{center}
\mbox{\epsfig{file=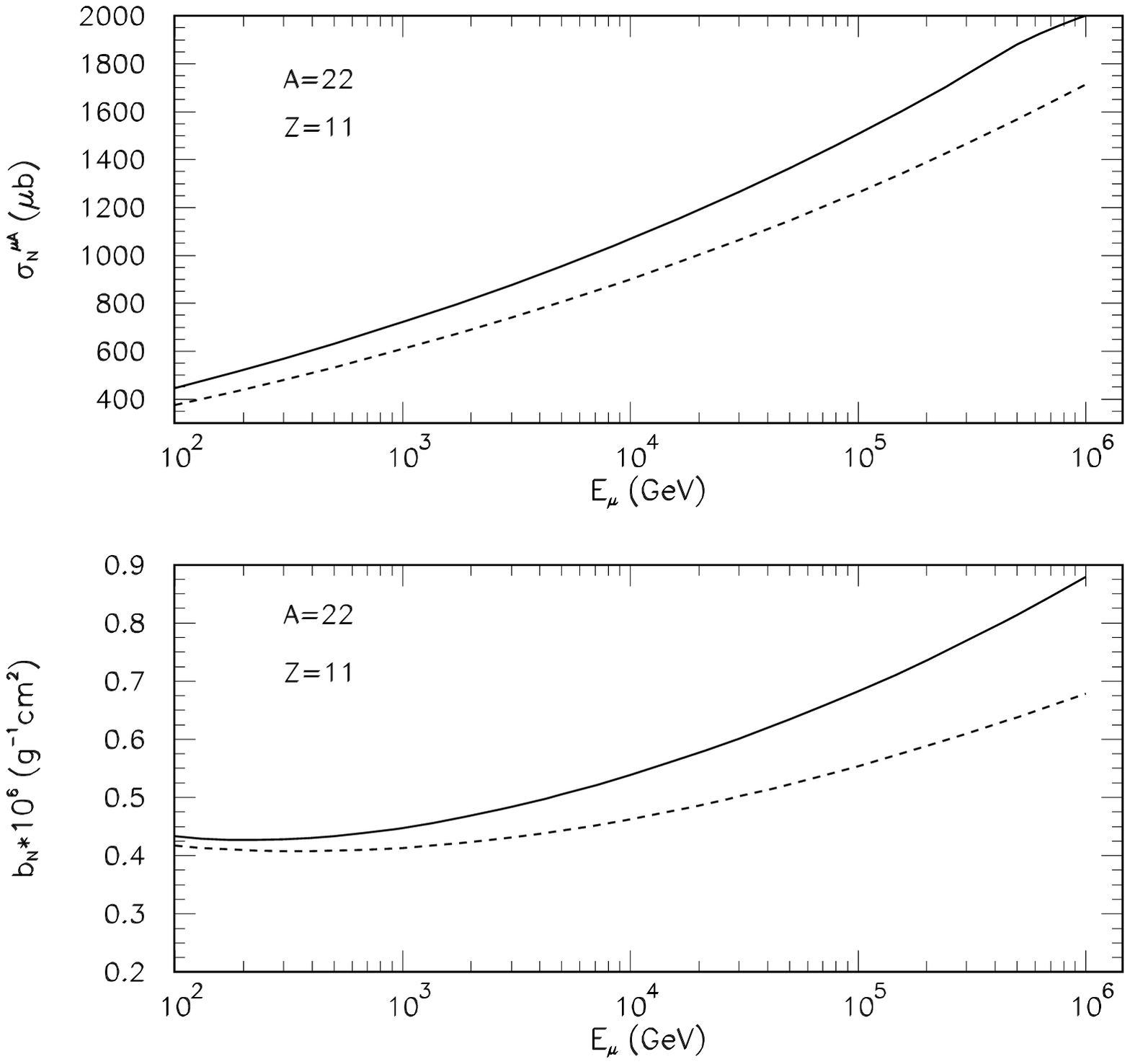,height=15cm,width=16cm}}
\vspace{2mm}
\noindent
\small
\end{center}
{\sf Fig.~15:}~The total cross-section $\sigma_{\mu A}$ and muon energy
 loss $b_n$ for muon inelastic scattering in standard rock as a function
of muon energy $E_{\mu}$, as compared to those from Ref.[4]. 
\newpage
\begin{center}
\mbox{\epsfig{file=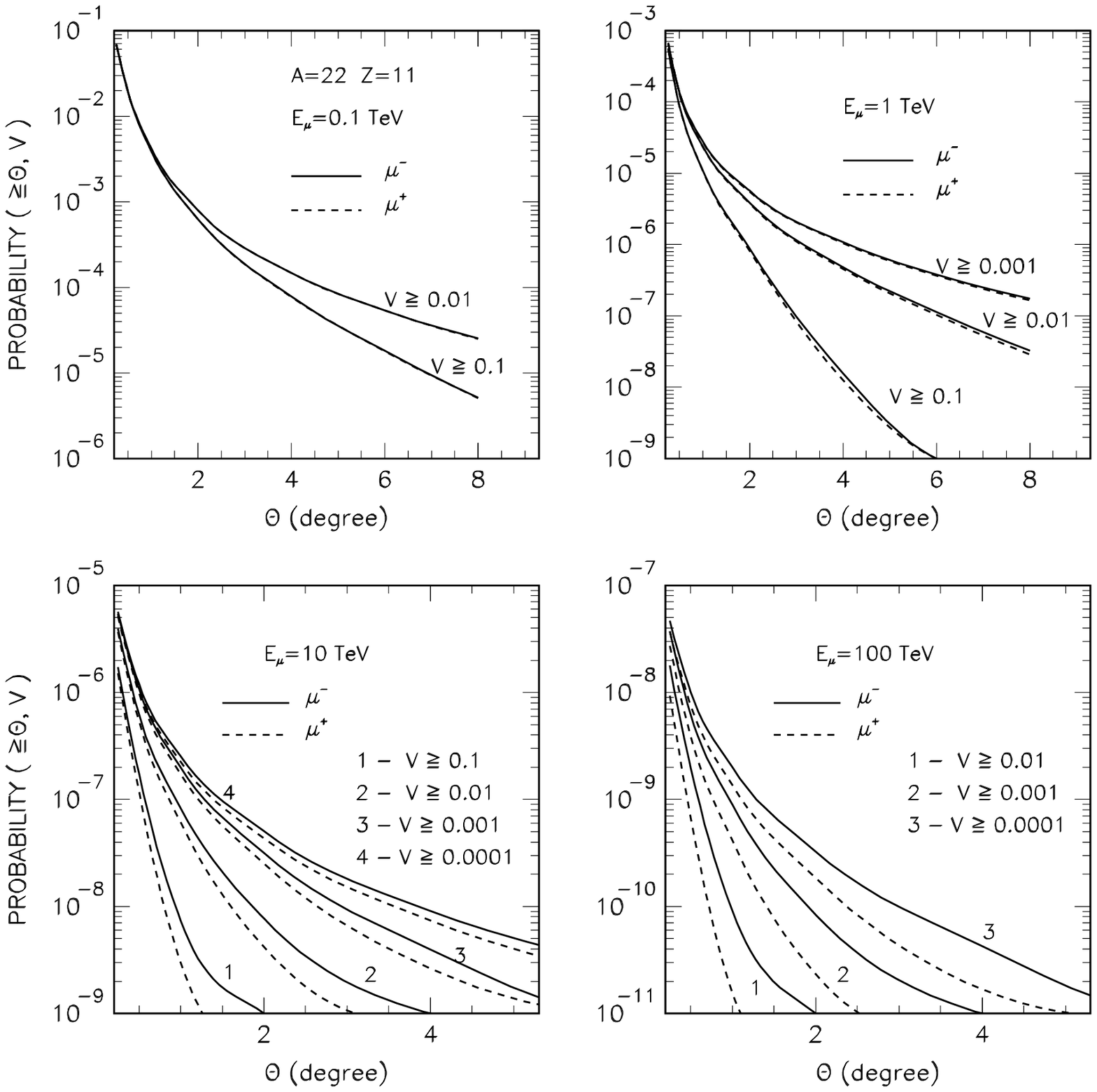,height=15cm,width=16cm}}
\vspace{2mm}
\noindent
\small
\end{center}
{\sf Fig.~16:}~ The probabilities $P(\ge\theta,\ge v)$ of muon scattering 
in single interaction at angle larger than $\theta$ with energy of 
outgoing muon $E'\ge vE$ as a function of $\theta$ for fixed values of $v$
and incoming  muon energies. Solid (dashed) lines are for positiv (negativ) 
muon charge.   

\normalsize
\end{document}